\documentclass[preprint]{elsarticle}
\usepackage{amsmath}
\usepackage{amssymb}
\usepackage{graphicx}
\usepackage{tikz}
\usepackage{sistyle}
\usepackage{color,colortbl}
\usepackage{caption}
\usepackage{subcaption}
\usepackage{ulem,cancel}
\usepackage[pdf]{pstricks}
\usepackage{epsfig,epsf,amsfonts,amscd}

\usepackage{amstext}

\journal{Physica D}
\biboptions{square,comma}

\begin{document}

\begin{frontmatter}

\title{Stationary fluxon properties 
in the Josephson junction ladder and their compact - non-compact
transitions}
\author[ap]{Andrii O. Prykhodko}
\ead{andriyprychodko@gmail.com}

\author[is]{Ivan O. Starodub}
\ead{starodub@bitp.kyiv.ua}

\author[yz]{Yaroslav Zolotaryuk\corref{cor1}}
\cortext[cor1]{Corresponding author}
\ead{yzolo@bitp.kyiv.ua}

%\address
%{Kyiv Academic University, 36 Vernadsky blvd., Kyiv 03142, Ukraine}
\address{
Bogolyubov Institute for Theoretical Physics, National Academy of
Sciences of Ukraine, Kyiv 03143, Ukraine}

\begin{abstract}
Linear stability of stationary fluxons in Josephson junction ladders
is investigated. The main focus is on the transition from
non-compact fluxons, where the phase distribution is
spread along the entire ladder, to \textit{compact} ones that
are localized on only a few junctions. Such compact fluxons cannot exist
in a standard parallel array of Josephson junctions, which is described by the
well-known discrete sine-Gordon equation (also known as the Frenkel--Kontorova
model). By performing a linear stability analysis of the
stationary fluxons, we identify the localized mode
responsible for transforming a non-compact fluxon into
a compact one. The critical anisotropy value at which this
transition takes place is computed analytically in the limit
of an infinite ladder.
\end{abstract}

\begin{keyword}
Josephson junction arrays \sep flat bands \sep
fluxons \sep stability \sep localized modes
\end{keyword}

\end{frontmatter}

%####################################################################
%####################################################################
\section{Introduction}
\label{intro}
%####################################################################
%####################################################################

Josephson junction arrays provide a tunable platform for studying how various collective effects emerge from coupled nonlinear oscillators.
Josephson junction ladders (JJLs) are known for a number of
nontrivial dynamical phenomena, such as the existence of
discrete breathers \cite{tmo00prl,baufz00prl}, flat 
bands in the plasmon spectrum \cite{fs99jpc,mffzp01pre,bz22jpc}, and
complex fluxon--breather interactions \cite{tmo02prb}, to name a few.
None of these phenomena occur in standard parallel arrays
of Josephson junctions \cite{wzso96pd}.

Fluxons are nonlinear topological excitations in Josephson-junction-based
systems. They are structurally stable and carry a magnetic
flux quantum. Fluxon properties in parallel arrays are 
well studied \cite{wzso96pd,u98pd}. Although stationary fluxons in JJLs 
have been investigated since the 1990s \cite{mff95prbII,gfpg96pla}, 
the role of ladder anisotropy---defined as the ratio of the critical 
currents in the ladder rungs and sides---requires further attention. 
Indeed, it was recently established \cite{psz25arx} that so-called 
\textit{compact} fluxons can exist in JJLs, whereas they are absent in parallel 
Josephson junction arrays. Compact fluxons exhibit an exact step-like
Josephson phase distribution of the form $\ldots, 0, 0, 0, 2\pi, 2\pi, \ldots$
and are exact solutions of the equations of motion.
While compact localized states are known in other
physical systems \cite{aks10prl,ssmjsvs20pra}, compact states in JJLs 
had not been reported until recently. 

The aim of this work is to examine the linear stability of
compact and non-compact stationary fluxons, investigate the
compactification (i.e., compact--non-compact) transition,
and analytically determine the existence diagram of non-compact
fluxons.

The paper is organized as follows. The model, equations of motion, 
and general formulation of the problem are presented in Section~\ref{model}. 
The main results are detailed in Section~\ref{sec3}, and our discussion
and conclusions are given in the final section.

%####################################################################
%####################################################################
\section{Model and formulation of the problem}
\label{model}
%####################################################################
%####################################################################

The linear JJL consists of $N+1$ vertical junctions, $N$ top and
$N$ bottom horizontal junctions, hence, the primitive cell of the
ladder consists of three junctions. The Josephson phases of individual
junctions are the differences of the wavefunction phases of the
superconducting leads. The state of the
ladder is defined by the vector $\mathbf{\Psi}=(\psi_1,\ldots,\psi_N,\psi_{N+1})^T$
where the vector $\psi_n$ describes the $n$th ladder cell and
 itself  consists of three components
$\psi_n=(\phi_n^{(v)}, \phi_{n,1}^{(h)}, \phi_{n,2}^{(h)})$. 
The superscripts $^{v,h}$ describe the vertical (rungs) or
horizontal (sides) of the ladder, and the subscript $k$
in $\phi_{n,k}^{(h)}$ corresponds to the bottom $k=1$ or top $k=2$
horizontal junctions.
The last component of $\mathbf{\Psi}$ is just a scalar
$\phi_{N+1}^{(v)}$.

The dynamics of the Josephson phases is governed by the
set of coupled discrete sine-Gordon equations that are the
consequence of the Josephson laws \cite{barone82}, Kirchhoff equations
and flux quantization in the ladder cell. 
The total energy of the linear ladder reads, normalized to the 
Josephson energy of an individual vertical junction $E_J=\hbar I_c^{(v)}/(2e)$ reads
%--------------------------------------------------------------------
\begin{eqnarray}\label{en}
&&H=\sum_{n=1}^N\frac{1}{2} \left [{\dot{\phi}_n^{(v)}}{}^2 +
\eta \sum_{k=1}^2 {\dot{\phi}_{n,k}^{(h)}}{}^2 \right ]+
\frac{1}{2} {\dot{\phi}_{N+1}^{(v)}}{}^2+\\
\nonumber
&&+\frac{1}{2\beta_L}\sum_{n=1}^{N+1}\left( \phi^{(v)}_{n+1}-\phi^{(v)}_{n}+
\phi^{(h)}_{n,2}-\phi^{(h)}_{n,1}-2\pi f\right)^2 +\\
\nonumber
&&+ \sum_{n=1}^{N+1} \left[1-\cos{\phi}^{(v)}_n -\gamma{\phi}^{(v)}_n \right]+ 
\eta \sum_{n=1}^{N}\sum_{k=1}^2 (1-\cos{\phi}^{(h)}_{n,k} ).
\nonumber
\end{eqnarray}
%--------------------------------------------------------------------
Here time is normalized to the inverse Josephson plasma frequency,
$t\to t\omega_p$, $\omega_p=\sqrt{2\pi I^{(v)}_c/(\Phi_0 C_v)}$
where $I^{(v)}_c$ is the critical current of vertical junctions,
$C_v$ is their capacitance and $\Phi_0=\pi\hbar /e$ is the
magnetic flux quantum. The frustration parameter $f=\Phi/\Phi_0$
is the ratio of the external magnetic flux $\Phi$ to the flux
quantum. Throughout the paper we will consider only the zero
flux $f=0$ situation. There also are two important dimensionless
physical parameters of the ladder:
%--------------------------------------------------------------------
\begin{equation}\label{b_e}
{\beta_L}= \frac{2\pi L I^{(v)}_c}{\Phi_0},\;
\eta=\frac{I^{(h)}_c}{I^{(v)}_c}=\frac{C_h}{C_v}=\frac{R_v}{R_h}, 
\end{equation}
%--------------------------------------------------------------------
Here $L$ is the self-inductance coefficient of the ladder cell, therefore
the parameter $\beta_L$ is the dimensionless self-inductance that 
describes the discreteness of the ladder. In other words, it 
is the ratio of the cell self-inductance to the Josephson inductance.
Another parameter is anisotropy $\eta$ that roughly measures the
proportion of energy concentrated in the subsystem of horizontal
and vertical junctions.
The ladder is additionally biased by the external constant current 
$\gamma={I_B}/{I^{(v)}_c}$ applied to each vertical junction.
In the final equations of motion we have to add the dissipative 
term $\alpha \dot\phi^{(v,h)}_{n}$, 
$\alpha=\hbar \omega_p/(2eI_v^{(v)}R_v)$ with $R_v$ being the 
vertical junction resistance. This term accounts for the normal electron flow across the
respective junction. The equations have been derived by the 
previous authors on several occasions\cite{fmamfm98pd,fs99jpc,baufz00prl,tmo02prb} and have the following form:
%-------------------------------2-5----------------------------------
\begin{eqnarray}\nonumber
\!\!\!\!\!\!\!\!\!\!\!\!\!\!\!\!\!\!\!\!
&&\ddot{\phi}^{(v)}_{n} + \alpha \dot{\phi}^{(v)}_{n}+\sin \phi^{(v)}_{n}=\gamma+\frac{1}{\beta_L} \left ( 
\phi^{(v)}_{n+1}-2\phi^{(v)}_{n}+\phi^{(v)}_{n-1}+\phi^{(h)}_{n,2}-\phi^{(h)}_{n,1}-\right.\\
\label{1}
\!\!\!\!\!\!\!\!\!\!\!\!\!\!\!\!\!\!\!\!&& \left.
-\phi^{(h)}_{n-1,2}+\phi^{(h)}_{n-1,1} \right ),\;\; n=\overline{2,N},\\
\!\!\!\!\!\!\!\!\!\!\!\!\!\!\!\!\!\!\!\!&& \ddot{\phi}^{(h)}_{n,k}+\alpha \dot{\phi}^{(h)}_{n,k}+\sin \phi^{(h)}_{n,k}=-
\frac{(-1)^k}{\eta\beta_L} \left( \phi^{(v)}_{n+1}-\phi^{(v)}_{n}+\phi^{(h)}_{n,2}-\phi^{(h)}_{n,1}\right ),\,n=\overline{1,N}  \label{2}\\
\!\!\!\!\!\!\!\!\!\!\!\!\!\!\!\!\!\!\!\!
&&\ddot{\phi}^{(v)}_1 + \alpha \dot{\phi}^{(v)}_1+
\sin \phi^{(v)}_1=\gamma+\frac{1}{\beta_L}
({\phi}^{(v)}_{2}-{\phi}^{(v)}_{1}+\phi^{(h)}_{1,2}-\phi^{(h)}_{1,1}),
\label{3}\\
\!\!\!\!\!\!\!\!\!\!\!\!\!\!\!\!\!\!\!\!&&\ddot{\phi}^{(v)}_{N+1} + \alpha \dot{\phi}^{(v)}_{N+1} +\sin \phi^{(v)}_{N+1}=\gamma-\frac{1}{\beta_L}
({\phi}^{(v)}_{N+1}-{\phi}^{(v)}_N+{\phi}^{(h)}_{N,2}-{\phi}^{(h)}_{N,1}), \label{4}
\end{eqnarray}
%--------------------------------------------------------------------
where Eqs. (\ref{3})-(\ref{4}) correspond to the boundary conditions on
the ladder edges. 

Linearization around the ground state (superconducting state) 
brings the following dispersion law \cite{baufz00prl,mffzp01pre}:
\begin{eqnarray}
\label{d1}
&&\omega(q)=\left \{
\begin{array}{cccc}
\omega_0(q)&=&&1,\\
&& \\
\omega_\pm(q)&=&&\sqrt{W_+(q)\pm
 \sqrt{W_-^2(q)+4\frac{1-\cos q}{\eta\beta_L^2}}},\\
&&& W_{\pm}(q)=\frac{1}{2}\left [1\pm \left(\omega^2_{||}(q)+\frac{2}{\eta\beta_L} \right) \right]
\end{array}
\right .\\
&&\omega_{||}(q)=\sqrt{1+\frac{4}{\beta_L}\sin^2 \frac{q}{2}},
\label{d2}
\end{eqnarray}
where $\omega_{||}(q)$ is the plasmon dispersion law for the
standard parallel array\cite{wzso96pd}. The dispersion law (\ref{d1})
has one completely flat band $\omega_0$, one strongly dispersive band $\omega_+$ and
one weakly dispersive band $\omega_-$. In the zero bias case ($\gamma=0$)
the weakly dispersive band coincides with the flat one. 

Fluxons are kinks that carry magnetic flux quantum. Hence, they
should connect two different ground states of the ladder,
$\phi^{v}=\arcsin\gamma$ and $\phi^{(v)}=2\pi +\arcsin\gamma$. 
For the multifluxons instead of $2\pi$ one has to put $2\pi Q$,
where $Q$ is topological charge or the number of fluxons.
The existence of the standard 
fluxons for which the phase distribution is spread along the ladder
and decays almost exponentially\footnote{It decays exactly
exponentially in the limit $N\to \infty$.} away from the fluxon
center is a well-established fact \cite{gfpg96pla,tmo02prb}.
We will refer to these fluxons as {\it non-compact} ones.

It was established in \cite{psz25arx} that apart from these non-compact
fluxons there are completely compact states that are
exact solution of the equations of motion (\ref{1})-(\ref{4})
and have exactly step-like phase distributions.
They can be classified as antisymmetric and asymmetric.
Suppose a compact fluxon with an arbitrary integer topological
charge is centered at the ladder cell j. Then the antisymmetric
fluxon with topological charge $Q\in \mathbf{Z}$ 
can be written explicitly: 
%--------------------------------------------------------------------
\begin{eqnarray}\!\!\!\!\!\!\!\!\!\!\!\!\!
&&\phi^{(v)}_n=\left\{ \begin{array}{cc}
\arcsin \gamma,& n\le j,\\
2\pi Q+\arcsin\gamma,& n > j 
\end{array} \right., \,n=\overline{1,N+1}, \label{vert1}\\
\!\!\!\!\!\!\!\!\!\!\!\!\!\!
&&\phi^{(h)}_{n,1}=-\phi^{(h)}_{n,2}=\pi Q \delta_{nj},\;\; n=\overline{1,N}.
\label{antsym1}
\end{eqnarray}
%--------------------------------------------------------------------
It is antisymmetric w.r.t to the top-bottom inversion:
$\phi^{(h)}_{n,1}=-\phi^{(h)}_{n,2}$. 
 
Another type of compact fluxons are asymmetric ones: 
%--------------------------------------------------------------------
\begin{eqnarray}
&&\phi^{(h)}_{n,1}=2\pi Q \delta_{nj},\; \phi^{(h)}_{n,2}=0,\; n=\overline{1,N},
\label{asym1}\\
&& \mbox{or} \nonumber \\
&&\phi^{(h)}_{n,1}=0,\; \phi^{(h)}_{n,2}=-2\pi Q \delta_{nj},\;  n=\overline{1,N}
\label{asym2}
\end{eqnarray}
%--------------------------------------------------------------------
These ones do not possess the top-bottom asymmetry. Combinations
of different fluxon types placed at arbitrary sites $j_1\neq j_2 
\neq \cdots \neq j_K$ are possible. 

If the stationary solution 
\begin{equation}
\mathbf{\Psi_0}=(\ldots; \Psi_n^{(v),0},\Psi_{n,1}^{(h),0},\Psi_{n,2}^{(h),0}; \ldots; 
\Psi_N^{(v),0},\Psi_{N,1}^{(h),0},\Psi_{N,2}^{(h),0}; \Psi_{N+1}^{(v),0})^T,
\end{equation}
%--------------------------------------------------------------------
 of Eqs. (\ref{1})-(\ref{4}) is known,
it is straightforward to linearize them around it as 
$\mathbf{\Psi}=\mathbf{\Psi_0}+\mathbf{A} e^{\mu t}$. As a result
one gets linear equation for the amplitudes $\mathbf{A}$ and the condition
one the solution existence:
\begin{equation}
{M} \mathbf{A}= \mu^2 \mathbf{A} \Rightarrow \det ({M}-\mu^2 \mathbb{I}_{3N+1})=0,
\; \lambda=\mu^2,
\end{equation}
where $\mathbb{I}_M$ is the identity matrix and the total $(3N+1)\times (3N+1)$ $M - \lambda \hat{\mathbb{I}}_{3N+1}$  matrix has the 
following form:
%------------------------------------7-------------------------------
\begin{equation}
M - \lambda \mathbb{I}_{3N+1} = \begin{pmatrix} 
D_1    & C   & 0   & \dots & 0    & 0\\
B      & D_2 & C   & \dots & 0    & 0\\
0      & B      & D_3    & \dots  &   0     & 0\\
\vdots & \vdots & \ddots & \ddots & \vdots  & \vdots \\
0      & 0      & \dots  & B & D_N & \mathbf{c} \\
0 & 0 & \dots & 0 & \mathbf{b}^T & {\cal D}_{3N+1}
\end{pmatrix}.
\label{7}
\end{equation}
%------------------------------------7-------------------------------
It has a tridiagonal block structure. The diagonal blocks are the $3\times 3$
matrices (except the last element, ${\cal D}_{3N+1}$, which is a scalar) have the 
following form:
%--------------------------------------------------------------------
\begin{eqnarray}
\!\!\!\!\!\!\!\!\! \nonumber
&&D_1=
%\left( \begin{array}{ccc}
%-\frac{1}{\beta_L}-\cos \Psi_1^{(v),0}-\lambda & \frac{1}{\beta_L \eta} & -%\frac{1}{\beta_L \eta} \\
%\frac{1}{\beta_L} & -\frac{1}{\beta_L\eta}-\cos \Psi_{1,1}^{(h),0}-\lambda & %\frac{1}{\beta_L \eta} \\
%-\frac{1}{\beta_L}& \frac{1}{\beta_L \eta} &-\frac{1}{\beta_L \eta}-\cos %\Psi_{1,2}^{(h),0}-\lambda 
%\end{array} \right)=  \\
%\!\!\!\!\!\!\!\!\! && = 
-\mbox{diag} \left (\cos \Psi_1^{(v),0},\cos \Psi_{1,1}^{(h),0},\cos \Psi_{1,2}^{(h),0}\right ) +
\frac{1}{\beta_L}\left( \begin{array}{ccc}
 -1 &  \frac{1}{\eta}  & -\frac{1}{\eta} \\
  1 & -\frac{1}{\eta}  & \frac{1}{\eta}  \\
 -1 &  \frac{1}{\eta}  & -\frac{1}{\eta}  
\end{array} \right)-\mathbb{I}_3\lambda,\\
\!\!\!\!\!\!\!\!\!
\nonumber
&&D_{n>1}=
%\left( \begin{array}{ccc}
%-\frac{2}{\beta_L}-\cos \Psi_n^{(v),0}-\lambda & \frac{1}{\beta_L \eta} & -\frac{1}{\beta_L \eta} \\
%\frac{1}{\beta_L} & -\frac{1}{\beta_L\eta}-\cos \Psi_{n,1}^{(h),0}-\lambda & \frac{1}{\beta_L \eta} \\
%-\frac{1}{\beta_L}& \frac{1}{\beta_L \eta} &-\frac{1}{\beta_L \eta}-\cos \Psi_{n,2}^{(h),0}-\lambda 
%\end{array} \right)=\\
%\!\!\!\!\!\!\!\!\! && =
 -\mbox{diag} \left (\cos \Psi_n^{(v),0},\cos \Psi_{n,1}^{(h),0},\cos \Psi_{n,2}^{(h),0}\right ) +
\frac{1}{\beta_L}\left( \begin{array}{ccc}
 -2 &  \frac{1}{\eta}  & -\frac{1}{\eta} \\
  1 & -\frac{1}{\eta}  & \frac{1}{\eta}  \\
 -1 &  \frac{1}{\eta}  & -\frac{1}{\eta}  
\end{array} \right)-\mathbb{I}_3\lambda,\\
\!\!\!\!\!\!\!\!\! 
&& {\cal D}_{3N+1}=-\frac{1}{\beta_L}-\cos \Psi_{N+1}^{(v),0}-\lambda.
%--------------------------------------------------------------------
\end{eqnarray}
The non-diagonal elements are given by the $3\times 3$ matrices $B$ and $C$
and vectors $\mathbf{b}$ and $\mathbf{c}$:
%--------------------------------------------------------------------
\begin{eqnarray}
\label{12}
B = \frac{1}{\beta_L}\begin{pmatrix} 
1 & -\frac{1}{\eta} & \frac{1}{\eta} \\ 
0 &   0       & 0 \\ 
0 &   0       & 0
\end{pmatrix},  \;\; C = \frac{1}{\beta_L}\begin{pmatrix} 
1  & 0 & 0 \\ 
-1 &  0 & 0 \\ 
1  &  0  & 0
\end{pmatrix},\\
 \mathbf{b}^T = \frac{1}{\beta_L}\left(1, -\frac{1}{\eta}, \frac{1}{\eta}\right), \;\; \mathbf{c}^T=\frac{1}{\beta_L}\left (1, -1, 1\right ).
 \label{13}
\end{eqnarray}
%--------------------------------------------------------------------

%####################################################################
%#################################################################### 
\section{Stationary fluxon stability}
\label{sec3}
%####################################################################
%####################################################################

\subsection{Compact fluxons}
In the case of stationary compact fluxons the profile $\mathbf{\Psi_0}$
can be written explicitly according to Eqs. (\ref{vert1})-(\ref{antsym1})
or Eqs. (\ref{asym1})-(\ref{asym2}). As a result the stability matrix simplifies 
greatly. Only the matrix diagonal contains the phase values, and,  
consequently, the $\cos \Phi^{(v),0}_{n}$ term attanis only the $\pm \sqrt{1-\gamma^2}$ values 
depending on the stationary fluxon type. Let us make a convention that
the Josephson phases at the edges of the ladder are always
in the ground (superconductive state), hence, the respective
first and the last diagonal blocks of $M-\lambda\mathbb{I}_{3N+1}$
can be written in the followin way:
%--------------------------------------------------------------------
\begin{eqnarray}
\!\!\!\!\!\!\!\!\! \label{17}
&&D_1=
%{\color{red}{
%\left( \begin{array}{ccc}
%-\frac{1}{\beta_L}-\sqrt{1-\gamma^2}-\lambda & \frac{1}{\beta_L \eta} & -%\frac{1}{\beta_L \eta} \\
%\frac{1}{\beta_L} & -\frac{1}{\beta_L\eta}-1-\lambda & \frac{1}{\beta_L \eta} \\
%-\frac{1}{\beta_L}& \frac{1}{\beta_L \eta} &-\frac{1}{\beta_L \eta}-1-\lambda 
%\end{array} \right)=}}  \\
%\!\!\!\!\!\!\!\!\! && = 
\frac{1}{\beta_L}\left( \begin{array}{ccc}
 -1 &  \frac{1}{\eta}  & -\frac{1}{\eta} \\
  1 & -\frac{1}{\eta}  & \frac{1}{\eta}  \\
 -1 &  \frac{1}{\eta}  & -\frac{1}{\eta}  
\end{array} \right)-\mbox{diag}(\sqrt{1-\gamma^2},1,1)-
\lambda\mathbb{I}_3,\\
\!\!\!\!\!\!\!\!\! 
&& {\cal D}_{3N+1}=-\frac{1}{\beta_L}-\sqrt{1-\gamma^2}-\lambda.
\label{18}
\end{eqnarray}
%--------------------------------------------------------------------
In the bulk of the ladder the matrix $D_n$ attains only two values,
$D_{\pm}$ depending whether the value of the horizontal phase
is $\pm \pi$ or $2\pi Q$:
%--------------------------------------------------------------------
\begin{eqnarray}
\!\!\!\!\!\!\!\!\! 
&&D_{\pm}=
%\left( \begin{array}{ccc}
%-\frac{2}{\beta_L}-\sqrt{1-\gamma^2}-\lambda & \frac{1}{\beta_L \eta} & -%\frac{1}{\beta_L \eta} \\
%\frac{1}{\beta_L} & -\frac{1}{\beta_L\eta}\mp 1-\lambda & \frac{1}{\beta_L \eta} \\
%-\frac{1}{\beta_L}& \frac{1}{\beta_L \eta} &-\frac{1}{\beta_L \eta}
%\mp 1-\lambda 
%\end{array} \right)=\\
%\!\!\!\!\!\!\!\!\! && = 
-\mbox{diag} \left (\sqrt{1-\gamma^2 },\pm 1, \pm 1 \right ) +
\frac{1}{\beta_L}\left( \begin{array}{ccc}
 -2 &  \frac{1}{\eta}  & -\frac{1}{\eta} \\
  1 & -\frac{1}{\eta}  & \frac{1}{\eta}  \\
 -1 &  \frac{1}{\eta}  & -\frac{1}{\eta}  
\end{array} \right)-\lambda\mathbb{I}_3.
\end{eqnarray}
%--------------------------------------------------------------------
Thus, we have $D_+$ if the respective phase is in the ground
state and $D_-$ if we have an antisymmetric fluxon centered there.
We can have $K$ fluxons centered at sites $j_1,j_2,\ldots,j_K$. 
The general charecteristic polynomial will be denoted as
\begin{equation}
\chi_{N;[j_1,\ldots,j_K]}(\lambda)=\det (M - \lambda \mathbb{I}_{3N+1}).
\end{equation}
Details of its evaluation are given in \ref{app1}.

The external bias enters only into the first diagonal element
of the matrices $D_n$. For the sake of simplicity we will 
consider only the zero-bias ($\gamma=0$) cases.
For the homogeneous unexcited ladder, all asymmetric fluxons and
antisymmetric fluxons with even topological charges the matrix
(\ref{7}) will have $D_n=D_+$ for $n=\overline{2,N}$.
The characteristic equation for the eigenvalues of the
matrix $M$ in that case can be reduced to 
\begin{eqnarray}
\!\!\!\!\!\!\!\!\!\!\!\!\!\!\!\!\!\!
&&   \chi_{N;0}(\lambda) = (-1)^{3N+1}(\lambda+1)\prod_{k=1}^N\left [ 
\lambda +\frac{2}{\beta_L}\left (1+\frac{1}{\eta} +
\cos \frac{\pi k}{N+1} \right )\right ].
\end{eqnarray}
%--------------------------------------------------------------------

Hence, we have $(3N+1)$-degenerate eigenvalue $\lambda=-1$ and
the band of $N$  eigenvalues
\begin{equation}
\lambda_k=-\frac{2}{\beta_L}\left (1+\frac{1}{\eta} +
\cos \frac{\pi k}{N+1}\right ).
\end{equation}
These eigenvalues
are always negative, and the ground state is stable as expected.
Since the respective vibration frequencies equal $\omega_k=\sqrt{-\lambda_k}$, it is not hard to see that they
correspond to the small-amplitude waves of the strongly dispersive
branch of the linear spectrum (\ref{d1}). The eigenvalue $\lambda=-1$
corresponds to the flat band.

For the antisymmetric compact fluxons with odd topological charges
or their combinations the situation is different. For the sites where 
the fluxon or fluxons are positioned we will have $\cos \pi$ terms
that will result in the appearance of the $D_-$ matrices in the
total stability matrix (\ref{7}). In this case it is not possible to
compute all the eigenvalues analytically. The characteristic
polynomial can be factorized only partially (see \ref{app1}). 

For the case of one antisymmetric fluxon positioned at the site
$j$ the characteristic polynomial has the following form:
%---------------------------20-21------------------------------------
\begin{eqnarray}
\!\!\!\!\!\!\!\!\!\!\!\!\!\!\!\!\!
&&\chi_{N;[j]}(\lambda) = (\lambda+1)^{2N-1}(\lambda-1) \left [ -(\lambda-1)P_N(z)
\label{20}
+\frac{4}{\eta\beta_L}P_{j-1}(z)P_{N-j}(z) \right ],\\
\!\!\!\!\!\!\!\!\!\!\!\!\!\!\!\!\!
&&z=-\lambda -1 -\frac{2}{\beta_L}\left (1+\frac{1}{\eta}\right),
\label{21}
\end{eqnarray}
%--------------------------------------------------------------------
where $P_n(t)=\frac{1}{\beta_L^n} U_n(\beta_L z/2)$ and $U_n$ are
Chebyshev polynomials of the second kind.
It has the $2N-1$-fold degenerate stable eigenvalue $\lambda=-1$ and
one unstable eigenvalue $\lambda=1$. Therefore, such configuration is
always unstable. The rest $N+1$ eigenvalues can be determined numerically
by finding the roots of the polynomial in the square brackets of
Eq. (\ref{20}).
In the case of $K$ fluxons the characteristic polynomial will
factorize as $(\lambda+1)^{2(N-K)-1}(\lambda-1)^K F_{N,K}(z)$.

\subsubsection{Small ladders}
It is instructive to look at small ladders with $N=1,2$ and $3$. In these
cases the analysis of the eigenvalues is simplified significantly.

\paragraph{One plaquette ($N=1$)}
Consider the single JJL plaquette or the ladder with $N=1$. We assume
no external bias ($\gamma=0$) and no applied magnetic field $f=0$.
There are four Josephson phases (two vertical and two horizontal)
and the equations of motion are
%--------------------------------------------------------------------
\begin{eqnarray}
&&\ddot{\phi}^{(v)}_n + \alpha \dot{\phi}^{(v)}_n+
\sin \phi^{(v)}_n=\gamma-\frac{(-1)^{n}\Upsilon}{\beta_L}, \; n=1,2,
\label{3a}\\
\!\!\!\!\!\!\!\!\!\!\!\!\!\!\!\!\!\!\!\!&&\ddot{\phi}^{(h)}_{k} + \alpha \dot{\phi}^{(h)}_{k} +\sin \phi^{(h)}_{k}=\gamma-\frac{(-1)^{k}}{\eta \beta_L} \Upsilon, \;k=1,2,
\label{3b}\\
&& {\Upsilon}= {\phi}^{(v)}_{2}-{\phi}^{(v)}_{1}+\phi^{(h)}_{2}-\phi^{(h)}_{1}.
\nonumber
\end{eqnarray}
%--------------------------------------------------------------------

It is not possible to distinguish between compact and non-compact fluxons
in this case. However, the asymmetric and antisymmetric stationary states
that break translational symmetry can exist:
\begin{eqnarray}
%--------------------------------------------------------------------
\mbox{Antisymmetric}:&&\; \phi^{(v)}_1=0, \phi^{(v)}_2=2\pi, \phi^{(h)}_1=-\pi,
\phi^{(h)}_2=\pi,\\
\mbox{Asymmetric}:&&\; \phi^{(v)}_1=0, \phi^{(v)}_2=2\pi, \phi^{(h)}_1=2\pi,
\phi^{(h)}_2=0.
\end{eqnarray}
In this case the stability problem reduces to calculating
the characteristic polynomial of
\begin{eqnarray}
\!\!\!\!\!\!\!\!\!\!
\chi_1(\lambda)= \det 
\begin{pmatrix} 
-\frac{1}{\beta_L}-1-\lambda & \frac{1}{\eta\beta_L} & -\frac{1}{\eta\beta_L} & \frac{1}{\beta_L} \\
\frac{1}{\beta_L} & -\frac{1}{\eta\beta_L} \pm 1-\lambda & \frac{1}{\eta\beta_L} &-\frac{1}{\beta_L} \\
-\frac{1}{\beta_L} & \frac{1}{\eta\beta_L} & -\frac{1}{\eta\beta_L} \pm 1-\lambda &\frac{1}{\beta_L} \\
\frac{1}{\beta_L} & -\frac{1}{\eta\beta_L} & \frac{1}{\eta\beta_L} & -\frac{1}{\beta_L}-1-\lambda 
\end{pmatrix}
\end{eqnarray}
The sign $+$ corresponds to the antisymmetric solution while $-$
corresponds to the asymmetric solution. 
The characteristic polynomial can be easily computed for both these cases.
For the asymmetric case it has the following form:
\begin{equation}
\chi_1(\lambda)=-(\lambda+1)^3 \left [\lambda+1+\frac{2(1+\eta)}{\beta_L\eta}\right ]=0.
\end{equation}
There is a triply degenerate eigenvalue $\lambda=-1$ and another one, $\lambda=-1-2/(\eta\beta_L)$, all of them being negative. That means that this state is stable
for all the model parameters.

In the antisymmetric case the eigenvalues are
\begin{equation}
\lambda_1=-1,\lambda_2=1, 
\lambda_{3,4}= \frac{-(1+\eta)\pm
\sqrt{(1+\eta)^2-\beta_L\eta [2-\eta(2+\beta_L)]}} {\beta_L\eta}.
\label{28}
\end{equation}
There exists a positive eigenvalue $\lambda=1$, hence, the antisymmetric
state is always unstable. The eigenvalue $\lambda_3$ (with "$-$" before
the square root) is always negative. Finally, the remaining eigenvalue,
$\lambda_4$ changes its sign from negative to positive at the critical 
value of anisotropy:
%------------------------------29------------------------------------
\begin{equation}
\eta_c=\frac{2}{2+\beta_L}. \label{29}
\end{equation}
%------------------------------29------------------------------------

\begin{figure}[htb]
\includegraphics[scale=0.3]{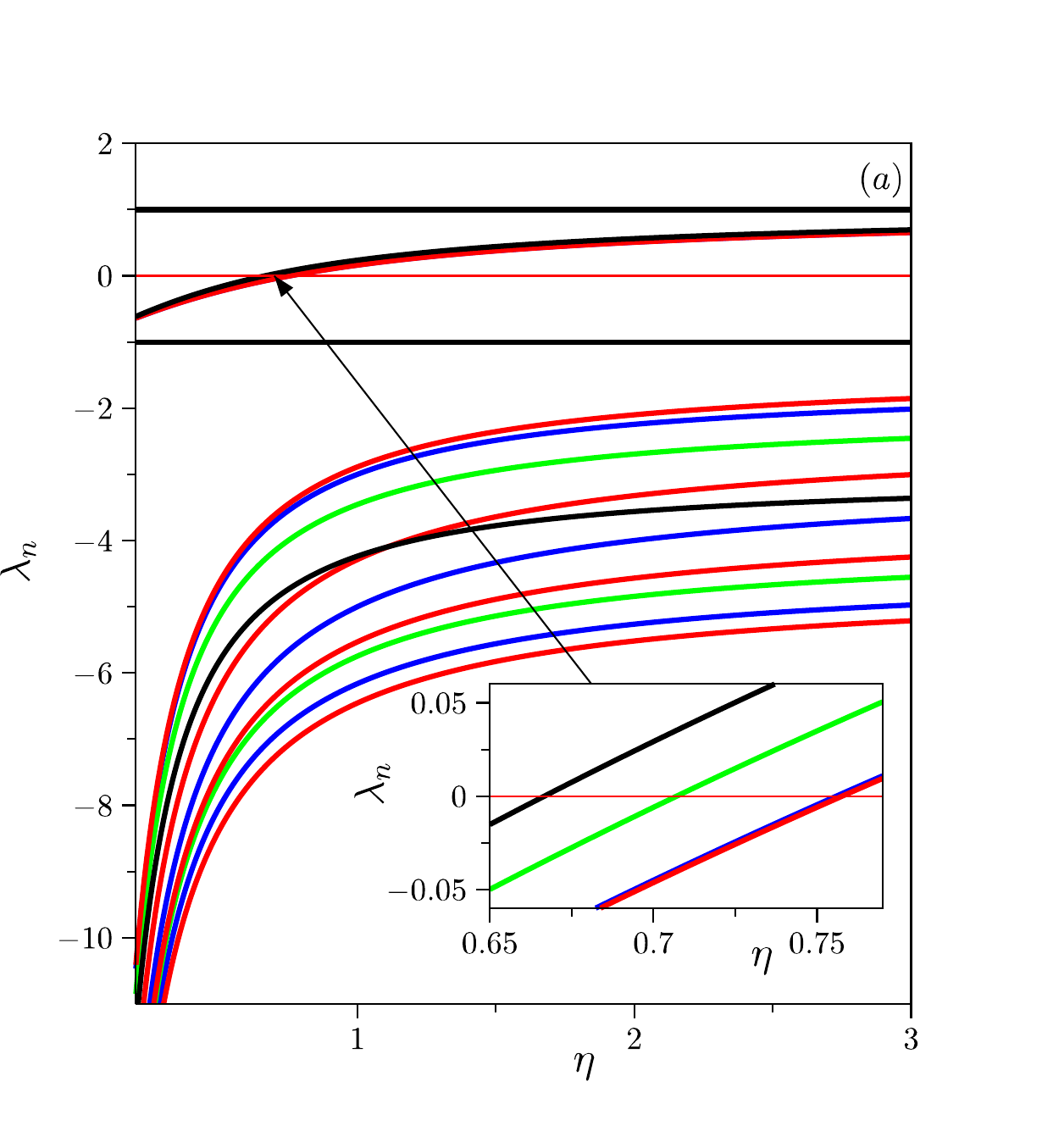}
\includegraphics[scale=0.3]{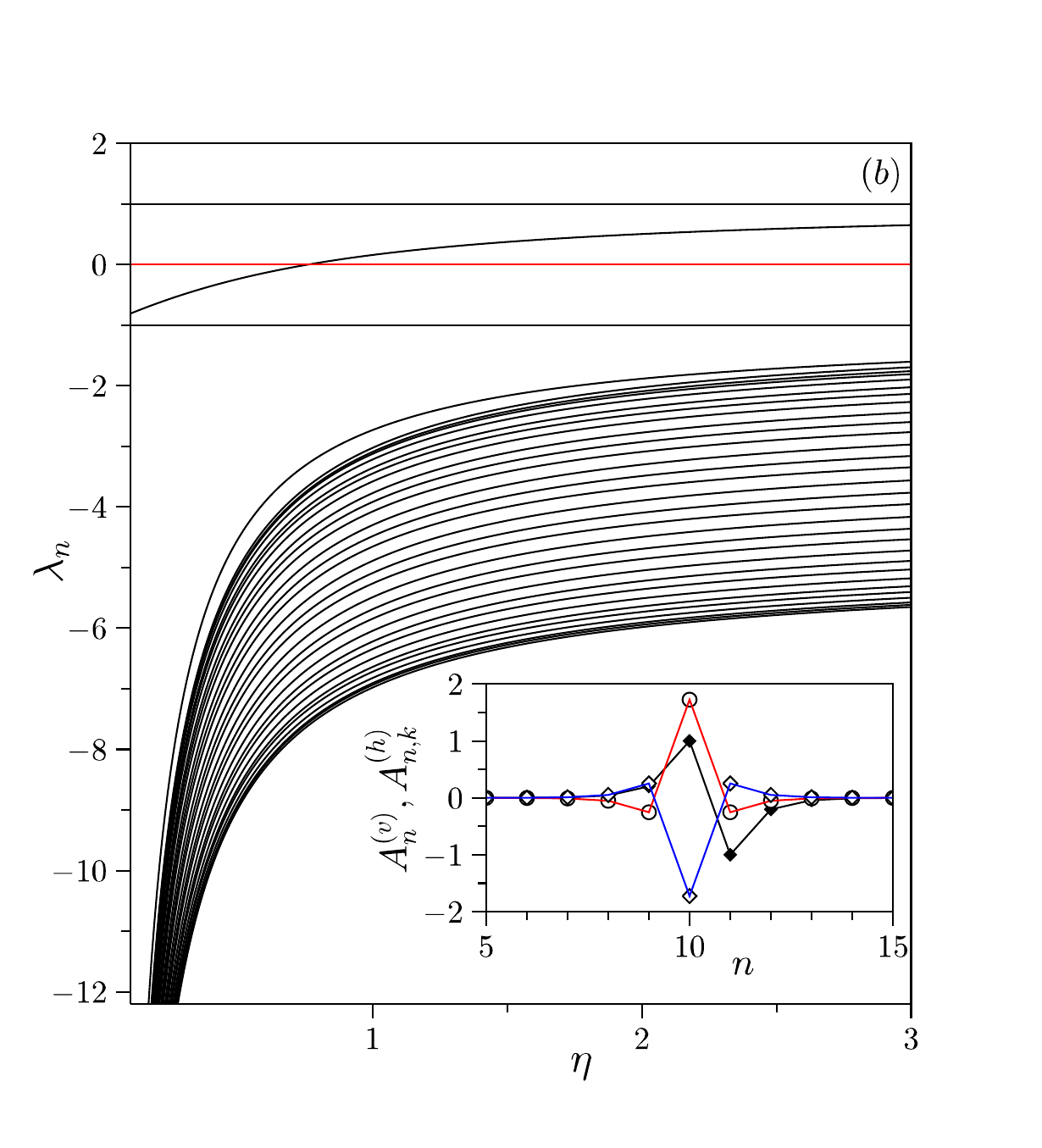}
\caption{(a)Eigenvalues as a function of anisotropy $\eta$ for 
ladders with one fluxon with $N=1$ (black), $N=2$ (green), $N=3$ (blue)
and $N=4$ (red). The thin red line marks the level $\lambda=0$.
The inset shows the details of the sign changing eigenvalue. \\
(b) Same for the $N=30$ ladder. The inset shows the components
of eigenvector $A_n^{(v)}$ (black, $\blacklozenge$), $A_{n,1}^{(h)}$ 
($\circ$, red) and $A_{n,2}^{(h)}$ ($\diamond$, blue) of the second largest eigenvalue at $\eta=1$.
}
\label{fig1}
\end{figure}

\paragraph{Cases $N=2,3$ and $4$}
In this case we have three vertical junctions and four horizontal ones.
The antisymmetric compact fluxon can be constructed by placing its center
at the site $j=2$. The total configuration will have the following
form:
\begin{equation}
\phi^{(v)}_1=\phi^{(v)}_2=0, \phi^{(v)}_3=2\pi, \\
\phi^{(h)}_{1,1}=\phi^{(h)}_{1,2}=0,\; \phi^{(h)}_{2,1}=-\pi, \phi^{(h)}_{2,2}=\pi.
\end{equation}
The characteristic polynomial allows for partial factorization. Using
the procedure described in the Appendix \ref{app1} and the general 
formal expression (\ref{20})-(\ref{21}) the characteristic
polynomial can be written explicitly:
\begin{eqnarray}
\chi_{2;[2]}(\lambda)=(\lambda+1)^3(\lambda-1)
\left \{ (1-\lambda)\left (z^2-\frac{1}{\beta_L^2}   \right) 
+\frac{4}{\eta \beta_L}  z \right\}. 
\end{eqnarray}
Apart from the eigenvalues $\lambda=1$ and $\lambda=-1$ which are
always present, there are three eigenvalues that are roots
of the cubic equation. In the case $N=3$, $j=2$ there are 
$10$ eigenvalues, the $5-$fold degenerate eigenvalue $\lambda=1$,
the unstable eigenvalue $\lambda=-1$. The $4$-order polynomial in
the square brackets of (\ref{20}) factorizes partially into 
a cubic polynomial and an additional negative eigenvalue 
$\lambda=-1-2(1+\eta)/(\eta \beta_L)$. In Fig. \ref{fig1}
all eigenvalues are plotted as a function of the anisotropy $\eta$.
%@@@@@@@@@@@@@@@@@@@@@@@@@@@@@@@@@@@@@@@@@@@@@@@@@@@@@@@@@@@@@@@@@@@@
%
%  Fig. (q1)
%____________________________________________________________________
%@@@@@@@@@@@@@@@@@@@@@@@@@@@@@@@@@@@@@@@@@@@@@@@@@@@@@@@@@@@@@@@@@@@@
Eigenvalues $\lambda=\pm 1$ that do not depend on the system parameters
are clearly seen. The eigenvalues that are associated with the 
linear waves have the property $\lim_{\eta\to 0}\lambda_n \to -\infty$ 
lie below the $\lambda=-1$ eigenvalue. There is one such eigenvalue
for $N=1$, two for $N=2$, three for $N=3$ and four for $N=4$. In all
of these cases there is one eigenvalue that lies in the limit 
$-1<\lambda<1$. Its exact value is given by Eq. (\ref{28}). Similarly,
there is one such eigenvalue for $N=2,3$ and $4$. The difference
between them is small, and they are almost undistinguishable
for $N=3$ and $4$ as shown in the inset of Fig. \ref{fig1}a.
The $\lambda=1$ eigenvalue
is the consequence of the flat band in the linear spectrum. 
The largest unstable eigenvalue $\lambda=1$ appears because the
horizontal phases $\phi_{j,1}^{(h)}=-\phi_{j,2}^{(h)}=\pi$ are
at the unstable equilibrium. The respective eigenvector 
has the following structure: $(0,1,1,0)^T$ for $N=1$, $(0,0,0,0,1,1,0)$
for $N=2$, $j=2$, and so on. That means that the unstable
direction of evolution is deviation of the $j$th horizontal
junctions.
The second largest eigenvalue
involves all junctions but is partially localized on the fluxon
center while the eigenvectors for the linear band are delocalized.

\subsubsection{General ladder size}

In Fig. \ref{fig1}b the stability eigenvalues are shown for the
large ladder of $N=3$. We observe the same patters with the
linear band moving to $-\infty$ as $\eta=0$, the degenerate
stable eigenvalue has the eigenvector $A^{(h)}_{n,k}=\delta_{nj}$,
$k=1,2$, $n=\overline{1,N}$ and $j$ is the fluxon position. 
The second largest
eigenvalue lies between the eigenvalues $-1$ and $1$. It is
spatially localized as shown in the inset of Fig. \ref{fig1}b
but it is not compact.
The critical value when the eigenvalue changes
its sign for $\beta_=1$ equals $\eta_c\approx 0.759$ which is
close to the $N=4$ shown in Fig. \ref{fig1}a.

\subsubsection{Limit $N\to \infty$ and localized mode.}

It is possible to compute the localized 
internal mode eigenvalue and
the critical anisotropy value when the eigenvalue changes
its sign. We use the same approach used for the impurity
modes calculation in solid state physics\cite{kittel}. 
Exponential localization of the eigenvector is assumed:
%--------------------------------------------------------------------
\begin{eqnarray}\label{32}
A_n^{(v)}&=& A_0 \left\{\begin{array}{cc} 
e^{\kappa (n-j)}, & n\le j ,\\ 
-e^{\kappa (j+1-n)}, & n \ge j+1,\end{array}\right. \\
A^{(h)}_{n,1}&=&-A^{(h)}_{n,2}.
\end{eqnarray}
%--------------------------------------------------------------------
Considering separately equations at the fluxon center (defect) 
site $n=j$:
$B A_{j-1}+ D_-A_j+C A_{j+1}=0$, 
[$A_n=(A_n^{(v)},A_{n,1}^{(h)},A_{n,2}^{(h)})^T$]
and away from the defect 
$B A_{n-1}+ D_+A_n+C A_{n+1}=0$ we can
express the eigenvector components of the horizontal junctions
through the vertical components:
%-------------------------------------------------------------------
\begin{eqnarray}
A^{(h)}_{n,1}=-\frac{\eta}{\eta \beta_L (\lambda-2\delta_{nj}+1)+2} (A_{n+1}^{(v)}-A_n^{(v)}), \label{34}
\end{eqnarray}
%-------------------------------------------------------------------
where the $\delta-$symbol in the denominator regulates whether
we are at the defect site or away from it. 
With the help Eq. (\ref{34}) the equations for the vertical 
components away from the fluxon ($n\neq j$) reduce to the following map:
\begin{eqnarray}\label{36}
A_{n+1}^{(v)}-\zeta A_{n}^{(v)}+A_{n-1}^{(v)}=0, 
\zeta=\beta_L(\lambda+1)+2\left (1+\frac{1}{\eta} \right).
\end{eqnarray}
For the fluxon center we have to rewrite the equation for
$A_{n+1}^{(v)}$'s taking into account that according to (\ref{34}) 
$A^{(h)}_{j,1}=-{\eta} 
(A_{j+1}^{(v)}-A_j^{(v)})/[\eta \beta_L (\lambda-1)+2]$ 
and $A^{(h)}_{j-1,1}=-{\eta}(A_{j}^{(v)}-A_{j-1}^{(v)})/[\eta \beta_L (\lambda+1)+2]$. From the 
ansatz (\ref{32}) and Eq. (\ref{36}) we obtain the decay exponent $\kappa$:
\begin{equation}
\cosh \kappa= \frac{\zeta}{2}. \label{37}
\end{equation}
From the same ansatz for the fluxon center we have  
$A_{j+1}^{(v)}=- A_{j}^{(v)}$, $A_{j-1}^{(v)}=e^{-\kappa} A_j^{(v)}$. Substituting
this into the main equation we obtain
\begin{equation}
e^{-\kappa}-3- \beta_L(\lambda+1)+2\left[ 
\frac{2}{\beta_L\eta (\lambda-1)+2}-\frac{e^{-\kappa}-1}{\beta_L\eta 
(\lambda+1)+2} \right]=0.
\label{38}
\end{equation} 
The decay exponent $\kappa$ is substituted from Eq. (\ref{37})
and, as a result, the final equation for the localized mode:
\begin{eqnarray}
%&&1+\zeta_-+\frac{1}{1+\zeta_-}=\zeta_+ \Rightarrow 
\nonumber
&& (\lambda+1)^2 \left \{(\lambda-1)^2+4\left [\frac{\eta+1}{\beta_L\eta}
(\lambda-1) 
+\frac{2\eta+1}{(\beta_L\eta)^2} \right]\right\}-\\
&&-\frac{16}{(\beta_L\eta)^2} \left [ (\eta+1)(\lambda+1)-1\right ]=0.
\label{39}
\end{eqnarray}
It is possible to find the roots of this equation numerically.
For the particular case of $\beta_L=1$ we observe
very high level of coincidence with the localized mode shown
in Fig. \ref{q1s2}.
To find the critical value of $\eta=\eta_c$ when the eigenvalue changes its
sign, we substitute $\lambda=0$ into Eq. (\ref{37}) for $\zeta$  
and into Eq. (\ref{38}). As a result we have the
square equation for $\eta_c$:
\begin{equation}
\eta^2+\frac{4}{\beta_L}\frac{\beta_L+2}{\beta_L+4}
-\frac{12}{\beta_L(\beta_L+4)}=0,
\end{equation}
which has a positive root:
\begin{equation}
\eta_c=2 \frac{2\sqrt{\beta_L^2+4\beta_L+1}-\beta_L-2}{\beta_L(\beta_L+4)}.
\label{41}
\end{equation}
This dependence has the following asypmtotics: $\eta_c({\beta_L\to 0})\to \infty$ and \\
${\eta_c}({\beta_L\to \infty}) \propto 4/\beta_L \to \infty$.
It breaks the plane ($\beta_L,\eta$) into two parts but to understand
the physical implications of it we have to look at the non-compact
states.

%###################################################################
\subsection{Non-compact fluxons}
%###################################################################

Our next point of interest is transition from compact
to non-compact solution and stability of non-compact fluxons.
The exact solutions for the non-compact solutions $\mathbf{\Psi_0}$
is not known, therefore it should be found numerically. To
do that the Newton method was used for Eqs. (\ref{1})-(\ref{4}).
The stability analysis will be performed numerically as well.

%\subsection{Case of topological charge $Q=\pm 1$.}

It is generally known \cite{bk03} that the discrete sine-Gordon model
that describes the standard parallel Josephson junction 
array has two stationary
kink (fluxon) solutions. The first one is a 
junction centered states have $\phi_{j}=\pi$ for the central
junction and is unstable. The second is cell-centered state
where the kink center lies exactly between the junctions
$j$ and $j+1$: $\pi-\phi_{-n+j}=-\pi+\phi_{n+j+1}$,
$n=0,\pm 1, \pm 2, \ldots$. This configuration is stable.

\begin{figure}[htb]
\includegraphics[scale=0.35]{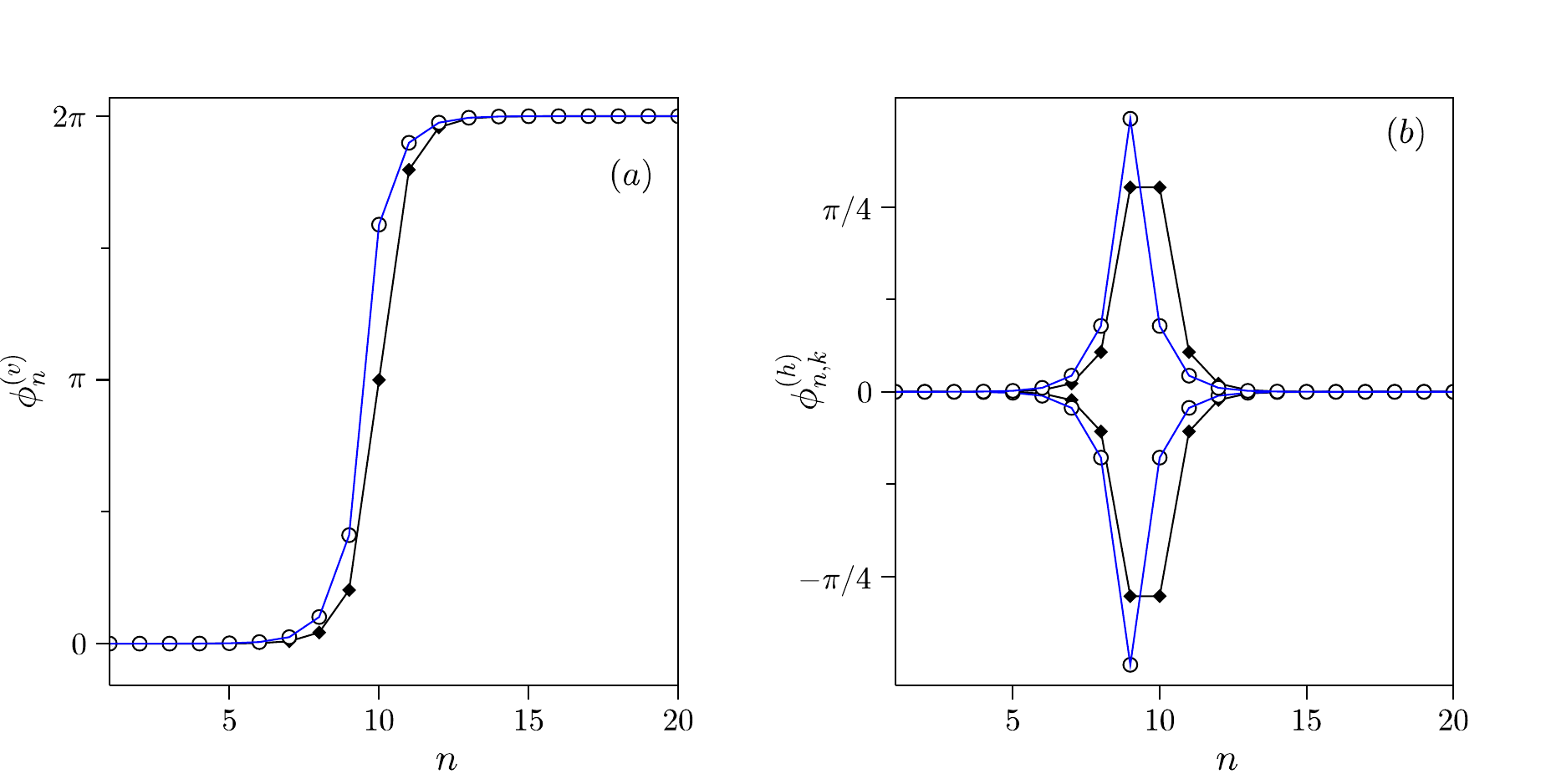}
\caption{Vertical (a) and horizontal (b) phase distribution for 
the junction-centered ($\blacklozenge$, black lines) 
and cell-centered ($\circ$, blue lines) stationary fluxons for 
$\beta_L=1$. The anisotropy equals $\eta=1.5$ for the cell-centered
fluxon and $\eta=1$ for the junction-centered one. 
The ladder size is $N=30$. 
}
\label{q1}
\end{figure}

In this work we observe similar situation for the JJL. 
The junction-centered
fluxon is always unstable, and the cell-centered one 
can be stable or unstable depending on the system parameters.
Examples of the junction-centered and cell-centered fluxons
are given in Fig. \ref{q1}. Note that for horizontal
distributions of the junction-centered profiles are cell-centered
and vice versa. 
%@@@@@@@@@@@@@@@@@@@@@@@@@@@@@@@@@@@@@@@@@@@@@@@@@@@@@@@@@@@@@@@@@@@@
%
%  Fig. (q1)
%____________________________________________________________________
%@@@@@@@@@@@@@@@@@@@@@@@@@@@@@@@@@@@@@@@@@@@@@@@@@@@@@@@@@@@@@@@@@@@@
The top-bottom antisymmetry is valid for both types of solutions.
Horizontal distributions $\phi^{(h)}_{n,1}$ are positive and
$\phi^{(h)}_{n,2}$ are negative.

\begin{figure}[htb]
\includegraphics[scale=0.3]{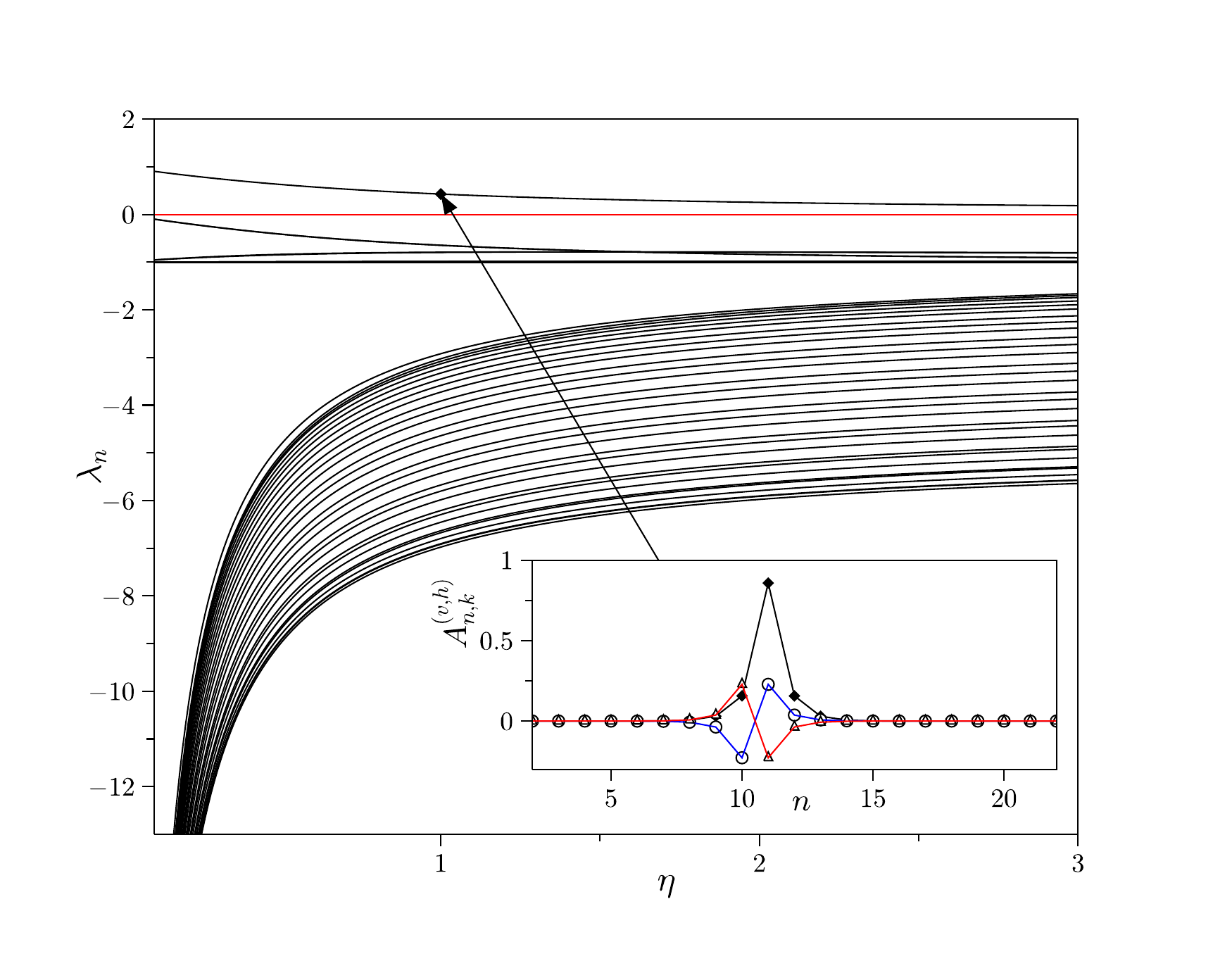}
\caption{Eigenvalues for the $Q=1$ junction-centeted kink with $\beta_L=1$ as a function
of the ladder anisotropy. The solid red line corresponds to the $\lambda=0$
value. The ladder size is $N=30$.
The inset shows the shape of the eigenvector
[$A_n^{(v)}$ (black, $\blacklozenge$), $A_{n,1}^{(h)}$ 
($\circ$, red) and $A_{n,2}^{(h)}$ ($\diamond$, blue)] 
 that corresponds to
the unstable eigenvalue at $\eta=1$ (marked by the arrow).
}
\label{q1s1}
\end{figure}

The eigenvalues of the stability matrix for the 
junction centered state as a function of the anisotropy parameter
are given in Fig. \ref{q1s1}. This figure clearly
demonstrates that junction-centered fluxons are unstable all the 
way.
%@@@@@@@@@@@@@@@@@@@@@@@@@@@@@@@@@@@@@@@@@@@@@@@@@@@@@@@@@@@@@@@@@@@@
%
%  Fig. (q1s1)
%____________________________________________________________________
%@@@@@@@@@@@@@@@@@@@@@@@@@@@@@@@@@@@@@@@@@@@@@@@@@@@@@@@@@@@@@@@@@@@@
There is only one positive eigenvalue and the respective
eigenvector is spatially localized. The structure of the 
unstable eigenvector is such that it pushes the central 
vertical junction from its unstable position $\phi^{(v)}_j=\pi$.
If perturbed, the junction-centered fluxon will transform
into the cell-centered one.

\begin{figure}[htb]
\includegraphics[scale=0.27]{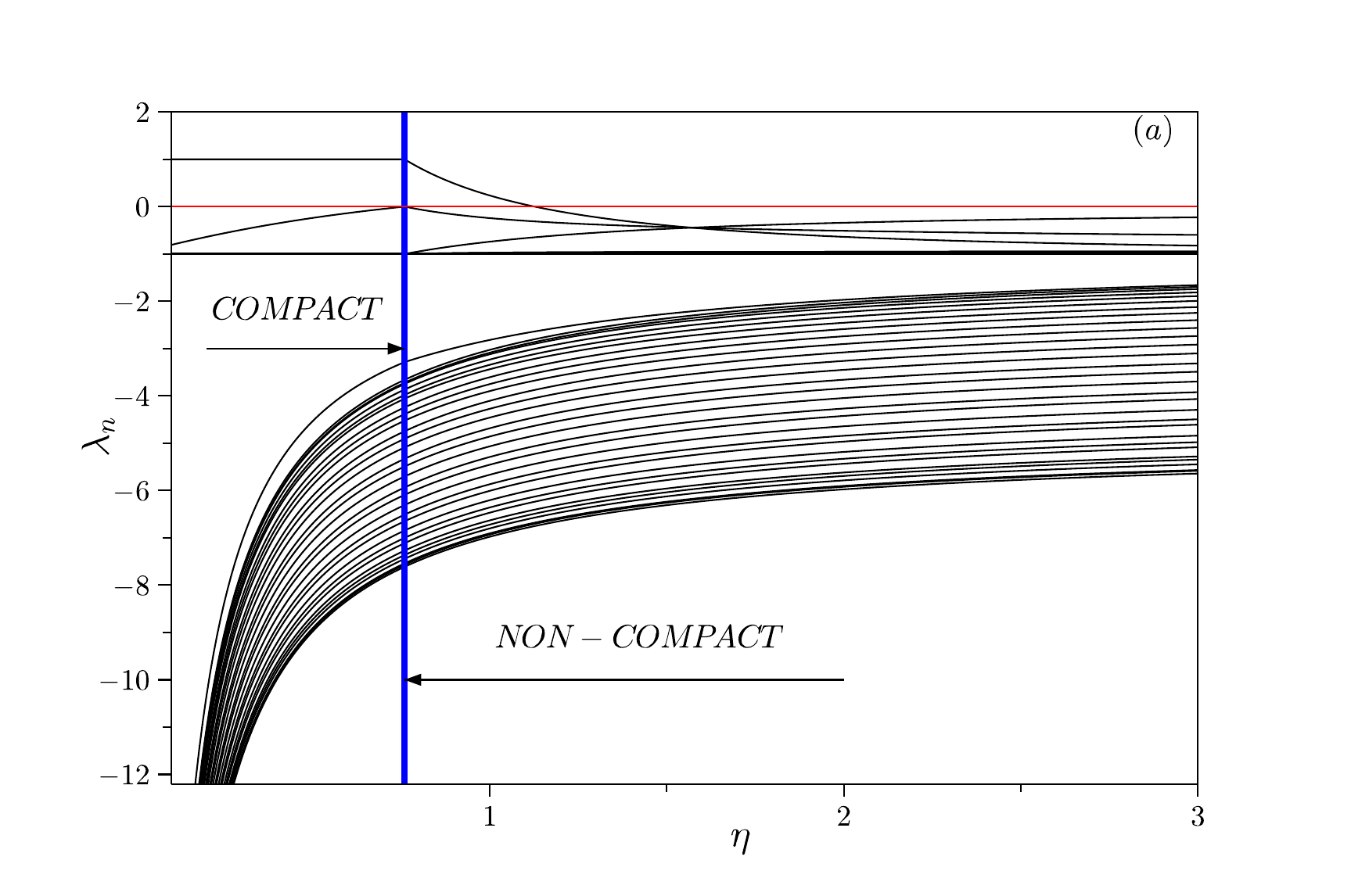}
\includegraphics[scale=0.28]{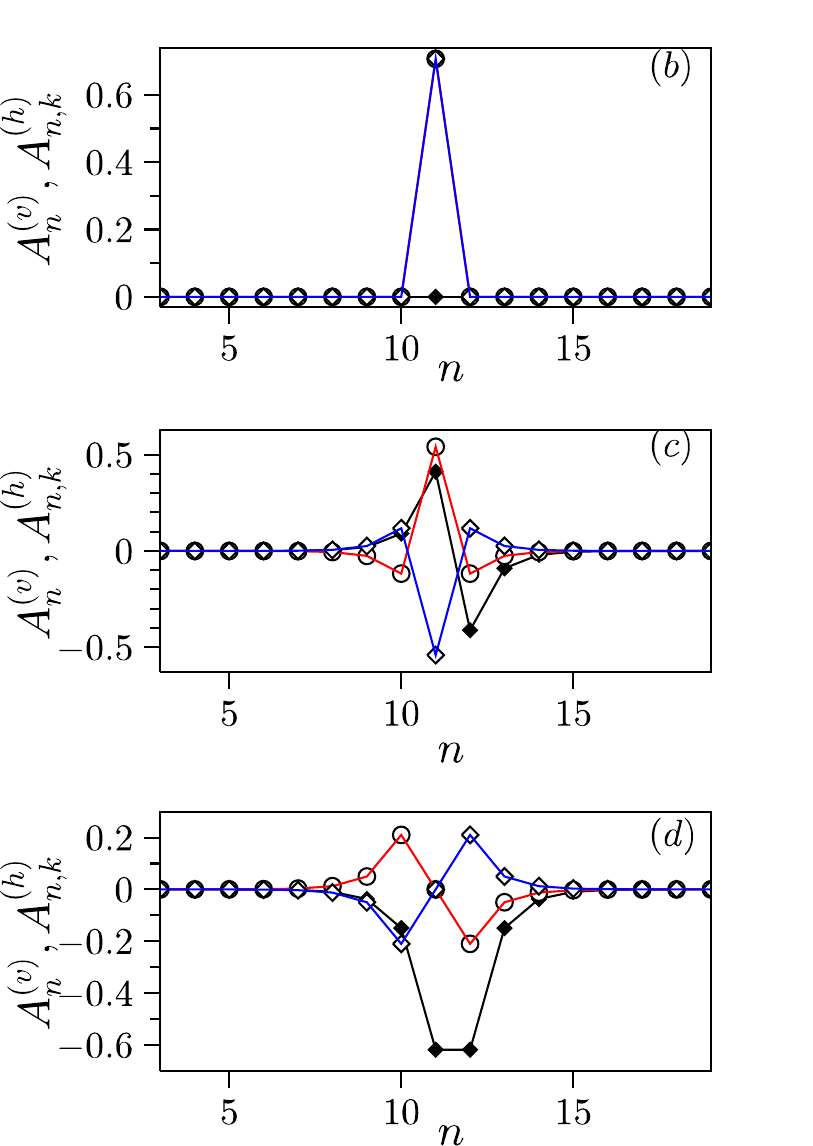}
\caption{(a) Eigenvalues for the $Q=1$ non-compact 
cell-centered fluxon and compact antisymmetic fluxons for $\beta_L=1$ as a function
of the ladder anisotropy. The solid red line corresponds to the $\lambda=0$
value. The ladder size is $N=30$.
Blue vertical line separates compact and non-compact cases.
(b)-(d) Eigenvectors that correspond the largest (b), second (c)
and third largest (d) eigenvalues for $\eta=1$. Markers correspond
to the same components as in previous figures.
}
\label{q1s2}
\end{figure}

The cell-centered fluxons are stable if the anisotropy $\eta$ is large 
enough. The eigenvalues as a function of $\eta$ are given
in Fig. \ref{q1s2}a. Here we observe the loss of stability
at some value $\eta\approx 1.18$ when the largest eigenvalue
crosses the $\lambda=0$ level and increases further. 
The respective unstable eigenvector [see Fig. \ref{q1s2}b] 
is compact and has
the same structure as the unstable eigenvector of the
antisymmetric compact fluxon: $A_n^{(v)}=0$, $A^{(h)}_{n,k}=\delta_{nj}$.
Extra 
decrease of $\eta$ brings a new instablity at $\eta\approx 0.759$
which appears to be fatal and the non-compact fluxon transforms
into the antisymmetric compact fluxon. The thick blue line separates
the eigenvalues for the non-compact fluxon (to the right) 
%@@@@@@@@@@@@@@@@@@@@@@@@@@@@@@@@@@@@@@@@@@@@@@@@@@@@@@@@@@@@@@@@@@@@
%
%  Fig. (q1s2)
%____________________________________________________________________
%@@@@@@@@@@@@@@@@@@@@@@@@@@@@@@@@@@@@@@@@@@@@@@@@@@@@@@@@@@@@@@@@@@@@
and for the compact antisymmetric (to the left) fluxons.
The dependencies $\lambda_n(\eta)$ in the interval $0<\eta<0.759$
are exactly the same as in Fig. \ref{fig1}b. 
The largest unstable eigenvalue becomes equal $1$ at the 
point of the non-compact fluxon disappearence.

So the second largest eigenvalue becomes zero at the
value $\eta_c\approx 0.759$ which coincides with the value of $\eta_c$
where the sign change
of the localized eigenvalue for the antisymmetric compact
fluxon (see Fig. \ref{fig1}b). The respective eigenvector has
the same spatial structure as the localized mode on the
compact antisymmetric fluxon as shown by the inset of Fig. \ref{fig1}b.
Now we fully understand the nature of the
localized mode on the compact state. If one takes the compact
antisymmetric fluxon and gradually increase $\eta$, the localized mode
will grow and, at some point it will be strong enough to cause
intstability. On the other hand, the non-compact fluxon becomes
narrower as $\eta_c$ is approached from above until it compactifies.

\begin{figure}[htb]
\includegraphics[scale=0.4]{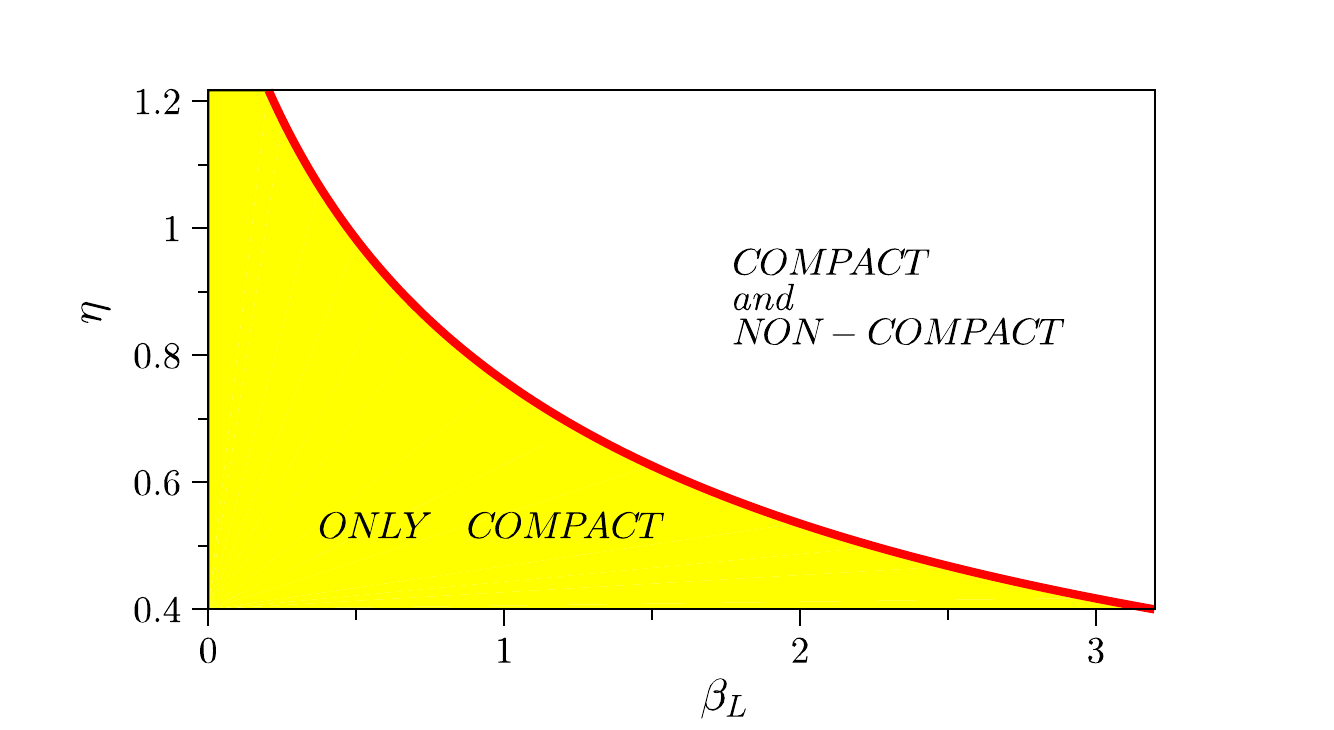}
\caption{Existence diagram of compact and non-compact stationary
fluxons on the inductance-anisotropy ($\beta_L-\eta$) parameter plane. In the white sector both compact and non-compact fluxons
exist. In the yellow sector compact and non-compact fluxons coexist.
The ladder size is $N=30.$
}
\label{ex}
\end{figure}

Finally, the numerically computed existence diagram for the
non-compact fluxons is compared with the analytically obtained
critical anisotropy $\eta_c$ Eq. (\ref{41}). We have simulated
numerically Eqs. (\ref{1})-(\ref{4}) using the 4th order Runge-Kutta method.
In order to provide convergence to the fluxon attractor state
we have introduced the damping $\alpha=0.2$. Since we are dealing
only with the stationary states, the actual role of damping is
just to provide attraction to the stable state. Therefore we do
not concern with the physically meaningful values of this parameter.
We start with the cell-centered non-compact fluxons at large
values of $\eta$ and path-follow these solutions until they
become exactly compact. Then we change $\beta_L$ and repeat the
procedure. The numerically calculated existence sector
is shown by white color in Fig. \ref{ex}. 
%@@@@@@@@@@@@@@@@@@@@@@@@@@@@@@@@@@@@@@@@@@@@@@@@@@@@@@@@@@@@@@@@@@@@
%
%  Fig. (ex)
%____________________________________________________________________
%@@@@@@@@@@@@@@@@@@@@@@@@@@@@@@@@@@@@@@@@@@@@@@@@@@@@@@@@@@@@@@@@@@@@
In the yellow sector we observe only stable compact fluxons. The
red line is analytically calculated critical anisotropy $\eta_c$
 Eq. (\ref{41}). We observe perfect coincidence of analytical
 and numerical results. Therefore, the bifurcation at
 $\eta=\eta_c$ signals disappearence of the non-compact fluxon.

It should be noted that the obtained existence diagram concerns
only the cell-centered fluxons because the junction-centered are
unstable for all $\eta$ and $\beta_L$. Also, the $\eta_c(\beta_L)$ 
   curve separates the {\it unstable} cell-centered fluxon from
   the compact one because the destabilizing bifurcation happens
   for larger values of $\eta$, see Fig. \ref{q1s2}a. Still, it
   is possible to continue the non-compact solution 
   into the unstable area using standard Runge-Kutta simulations. 
We have also noticed that not all perturbations destroy the 
unstable cell-centered fluxon. It is not destroyed if the
only the vertical system is perturbed. However, perturbation 
of the horizontal junctions leads to its transtion into the
compact asymmetric fluxon, which is always stable.

%\subsection{Case of topological charge $Q=\pm 2$.}
%
%Stationary fluxons with topological charge $Q=2$ can
%be considered as bound states of two fluxons with some
%separations between them. 
%{\color{red}{This issue for the DSG equation has been discussed
%in detail by \cite{bck00pd}}}
%
%
%\begin{figure}[htb]
%\includegraphics[scale=0.3]{/home/yzolo/WORK/JJA/JJL/q2s1.pdf}
%\includegraphics[scale=0.2]{/home/yzolo/WORK/JJA/JJL/evecs1.pdf}
%\caption{Eigenvalues for the $Q=2$ kink with $\beta_L=2$ as a function
%of the ladder anisotropy. The solid red line corresponds to the $\lambda=0$
%value. The ladder size is $N=30$.
%}
%\label{figx}
%\end{figure}

%####################################################################
%####################################################################
\section{Conclusions and discussion}
%####################################################################
%####################################################################

This work is primarily devoted to the linear stability of fluxons near the
transition between compact and non-compact profiles. 
The existence of compact fluxons, established in \cite{psz25arx},
is a direct consequence of the flat band in the linear spectrum of the ladder;
no such compact fluxons can exist in a standard parallel Josephson junction
array. 

Our results highlight the crucial role of the anisotropy
parameter $\eta$, defined as the ratio of the horizontal to the vertical 
junction critical currents. We found a critical anisotropy--inductance 
curve $\eta_c(\beta_L)$, which is a monotonically decreasing function of $\beta_L$. 
This curve divides the $\eta\text{--}\beta_L$
parameter plane into two regions: (i) for $\eta < \eta_c$, only compact states exist,
and (ii) for $\eta > \eta_c$, both compact and non-compact fluxons
coexist. We derived the critical function $\eta_c(\beta_L)$ 
analytically in the infinite-ladder limit. 

The mechanism underlying the compact--non-compact transition is governed by
a localized internal mode supported by the fluxon. Starting
from an antisymmetric compact fluxon at small values of $\eta$,
this localized (and spatially extended-tail) mode triggers a bifurcation
at $\eta = \eta_c$ that gives rise to the non-compact fluxon.
Conversely, when decreasing $\eta$ from large values toward smaller
ones, the initially stable non-compact fluxon first loses stability. 
A further decrease in anisotropy causes the second-largest
eigenvalue to cross zero, at which point the non-compact fluxon transforms into
a compact state.

Finally, we note that ladder anisotropy strongly governs how
trapped magnetic flux is distributed across the array. A promising 
direction for future research is the investigation
of plasmon scattering by compact and non-compact fluxons.

%####################################################################
%####################################################################
\section*{Acknowledgements}
%####################################################################
%####################################################################
We would like to thank the Armed Forces of Ukraine for providing security
to perform this work.
All authors acknowledge support from the National
Research Foundation of Ukraine, grant (2023.03.0097)
"Electronic and transport properties of Dirac materials and
Josephson junctions". 

%####################################################################
%####################################################################
\appendix
\section{Details of the stability matrix diagonalization of
and charecteristic polynomial calculation}
\label{app1}
%####################################################################
%####################################################################

To diagonalize the block-tridiagonal matrix
%------------------------------------7-------------------------------
\begin{equation}
M - \lambda \mathbb{I}_{3N+1} = \begin{pmatrix} 
D_1    & C   & 0   & \dots & 0    & 0\\
B      & D_2 & C   & \dots & 0    & 0\\
0      & B      & D_3    & \dots  &   0     & 0\\
\vdots & \vdots & \ddots & \ddots & \vdots  & \vdots \\
0      & 0      & \dots  & B & D_N & \mathbf{c} \\
0 & 0 & \dots & 0 & \mathbf{b}^T & {\cal D}_{3N+1}
\end{pmatrix}.
\label{a1}
\end{equation}
%------------------------------------7-------------------------------
one can use the Schur complement. The Schur complement for
the $2\times 2$ block matrix $M=\left(\begin{array}{ccc}  A & B \\ C&D   \end{array} 
\right)$
yields $\det M= \det A ~ \det (D-CA^{-1}B)$. 
The diagonal blocks of (\ref{app1}), $D_n$, $n=\overline{2,N}$, 
can attain only two values,
$D_-$, if there is an antisymmetric fluxon centered there or $D_+$ if the respective
primitive cell is in the equilibrium position. We will refer to
the $D_-$ blocks as defects.
By applying it simulaneously starting from the top left block $D_1$ till
the last block which is a scalar ${\cal D}_{3N+1}$, we can write down the
whole characteristic polynomial.
Suppose that there are $K$ defects positioned at the cells
$j_1,j_2,\ldots, j_K$. Then the diagonal blocks will form the sequence
%--------------------------------------------------------------------
\begin{equation}
D_1D_+\cdots D_+D_-D_+\cdots D_+ D_- D_+ {\cal D}_{3N+1}~.
\end{equation}
Only two diagonal elements ${D_{n}}_{22}$ and ${D_{n}}_{33}$ of the matrix $D_n$ define whether the matrix is $D_+$ or $D_-$. For the sake of simplicity
we denote
%--------------------------------------------------------------------
\begin{equation}\label{a3}
{D_{n}}_{22}={D_n}_{33}=\xi_n-\frac{1}{\eta\beta_L},\;\; 
\xi_n=\mp 1-\lambda=2\delta_{nj}-1-\lambda.
\end{equation}
%--------------------------------------------------------------------
After the first step of the Schur decomposition one obtains
\begin{eqnarray}
\det(M - \lambda \mathbb{I}_{3N+1})=\det D_1 ~ \det [S_{3N-2} -\mbox{diag}(g_1,0,\ldots,0) ],
\end{eqnarray}
%--------------------------------------------------------------------
where $S_{3N-2}$ is the submatrix of $M - \lambda \mathbb{I}_{3N+1}$
obtained by removing first row and first column.  The next submatrix, obtained
in the similar way from $S_{3N-2}$ will be denoted as $S_{3N-5}$ an so on.
The subscript of the matrix denotes its dimension. The matrix which is
subtracted from $S_{3N-2}$ has only one diagonal element $g_1$, because, due
to the structure of matrices $B$ and $C$ given by Eq. (\ref{12}) 
one determines that    
\begin{equation}
BD^{-1}_1C=\mbox{diag} (g_1, 0, 0), \; g_1=-\frac{1}{\beta_L}\frac{\frac{1}{\beta_L}
+\frac{2}{\eta\beta_L}}{\lambda+1+\frac{1}{\beta_L} +\frac{2}{\eta\beta_L}}.
\end{equation}
%--------------------------------------------------------------------

Hence, the total determinant calculation can be rewritten as
%--------------------------------------------------------------------
\begin{eqnarray}
\!\!\!\!\!\!\!\!\!\!\!\!\!\!\!
\nonumber
&&\det(M - \lambda \mathbb{I}_{3N+1})= \det D_1 \det [D_2-\mbox{diag}(g_1,0,0)]\times \\
\nonumber
\!\!\!\!\!\!\!\!\!\!\!\!\!\!\!
&& \times \det [S_{3N-2}-B [D_2-\mbox{diag}(g_1,0,0)]^{-1} C ] = \\
\nonumber
\!\!\!\!\!\!\!\!\!\!\!\!\!\!\!
&& = \det D_1 \det [D_2-\mbox{diag}(g_1,0,0)] \det [D_3-\mbox{diag}(g_2,0,0)]\times \\
\nonumber
\!\!\!\!\!\!\!\!\!\!\!\!\!\!\!
&& \times \det [S_{3N-5}-B [D_3-\mbox{diag}(g_2,0,0)]^{-1} C ]=\\
\!\!\!\!\!\!\!\!\!\!\!\!\!\!\!
&& \!\!\!\!\!\!\!\!\!\!\!\!\!\!\! \cdots \cdots \cdots \cdots \nonumber \\
\!\!\!\!\!\!\!\!\!\!\!\!\!\!\!
&& =\left(\prod_{n=1}^{N} \det \tilde{D}_n  \right) 
(-\lambda-1-\beta_L^{-1}-g_N),\;n=
\overline{2,N}, \label{a9}
\end{eqnarray}
%--------------------------------------------------------------------
where the matrices $\tilde{D}_n$ are obtained by subtracting 
the scalar $g_n$ from the matrix $D_n$:
%--------------------------------------------------------------------
\begin{equation}
{\tilde D}_1=D_1,\; {\tilde D}_n=D_n-\mbox{diag}(g_{n-1},0,0), \;\;
g_n=\frac{a_n+b_n g_{n-1}}{d_n+c_n g_{n-1}},\; n=\overline{2,N}.
\label{a10}
\end{equation}
%--------------------------------------------------------------------
As one can see that $g_n$ can be expressed through $g_{n-1}$ 
as a M\"obius transform. The coefficients of the transform depend whether
the respective matrix $D_n$ is $D_+$ or $D_-$, therefore the coefficients 
$a_n,b_n,c_n$ and $d_n$ in Eq. (\ref{a10}) are defined by the respective
matrix $D_n$ and, must be equipped with the subscripts
${}_\pm$:
%--------------------------------------------------------------------
\begin{eqnarray}
\nonumber
&&a_\pm=\frac{\mp 1-\lambda}{\beta^2_L}\left [\mp 1-\lambda-\frac{2}{\eta} 
\left (\lambda+1 +\frac{1}{\beta_L}  \right) \right ], \;\;\; 
b_\pm=\frac{2(\lambda\pm 1)}{\eta\beta_L^2},\\
&&c_\pm=(\lambda \pm 1)\left [\frac{2}{\eta\beta_L^2} -
\left (\lambda+1+\frac{2}{\beta_L}\right )\left ( \lambda \pm 1 +\frac{2}{\eta\beta_L}\right)\right ],  \nonumber \\
&& d_\pm=-(\mp 1 -\lambda)
\left(\mp 1 -\lambda -\frac{2}{\eta\beta_L}\right).
\end{eqnarray}
%--------------------------------------------------------------------
The respective matrix $\tilde{D}_n$ is defined by the matrix $D_n$ and
$g_n$. Its determinant can be computed and expressed through $g_{n-1}$
and the elements of the M\"obius transform:
%--------------------------------------------------------------------
\begin{equation}
\det \tilde{D}_n=d_n g_{n-1}+c_n.
\label{a13}
\end{equation}
%--------------------------------------------------------------------
The M\"obius transform (\ref{a10}) can be rewritten using a transfer-matrix 
approach:
\begin{eqnarray}
\left (\begin{array}{c} {x}_n \\ {y}_n \end{array} \right) =
\left (\begin{array}{cc} b_n & a_n \\ d_n & c_n \end{array} \right)
\left (\begin{array}{c} {x}_{n-1} \\ {y}_{n-1} \end{array} \right)\equiv
T_n \left (\begin{array}{c} {x}_{n-1} \\ {y}_{n-1} \end{array} \right), 
\;\; g_n=\frac{x_n}{y_n}.
\end{eqnarray}
%--------------------------------------------------------------------
Taking into account this representantion and Eq. (\ref{a13}) one
can deduce that $y_n=d_n x_{n-1}+c_n y_{n-1}=q_{n-1}(d_n g_{n-1}+c_n)=
y_{n-1} \det \tilde{D}_n$. As a result, the product in Eq. (\ref{a9}) 
equals $\prod_{n=1}^{N} \det \tilde{D}_n=y_N$. As a result, the 
characteristic polynomial can be rewritten as
%--------------------------------------------------------------------
\begin{eqnarray}\nonumber
&&\det(M - \lambda \mathbb{I}_{3N+1})=y_N \left (-1-\lambda-\frac{1}{\beta_L} -g_N \right )= \left (-1-\lambda-\frac{1}{\beta_L} \right )y_N-x_N=\\
&&=-\left (1, \lambda+1+\frac{1}{\beta_L}\right )^T \prod_{n=N}^1T_n 
\left (\begin{array}{c} x_1 \\ y_1 \end{array}\right ).
\label{a11}
\end{eqnarray}
%--------------------------------------------------------------------
We have rewritten the product of determinants as a product of
transfer matrices $T_n$. The matrix $T_n$ is defined by the coefficients
$a,b,c$ and $d$ and equals $T_+$ or $T_-$ depending on the position of
the defect in the ladder. This matrix can be decomposed as
%--------------------------------------------------------------------
\begin{eqnarray}
&&T_n=\xi_n^2 {G} {\cal T}_n G^{-1}, \;\;
G=\left (\begin{array}{cc} \frac{1}{\beta_L} & -\lambda-1-\frac{1}{\beta_L} \\ 
-1 & 1 \end{array} \right),\\
&&G^{-1}=-\frac{1}{\lambda+1} \left (\begin{array}{cc} 1&  
\lambda+1+\frac{1}{\beta_L} \\ 
 1 & \frac{1}{\beta_L} \end{array} \right),\\
&& {\cal T}_n= \left (\begin{array}{cc} 
-\lambda -1 -\frac{2}{\beta_L} \left (1-\frac{\lambda+1}{\eta\xi_n} \right ) & \frac{1}{\beta_L} \\
 \frac{1}{\beta_L} &0
\end{array} \right),
\end{eqnarray}
%--------------------------------------------------------------------
where $\xi_n$ is defined in Eq. (\ref{a3}). As a result, the expression for
the characteristic polynomial (\ref{a11}) can be rewritten in more
compact form:
%--------------------------------------------------------------------
\begin{eqnarray}
\det(M - \lambda \mathbb{I}_{3N+1})=-(\lambda+1) \left ( 
\prod_{n=1}^N \xi_n^2 \right)~ (1,0)^T \prod_{n=N}^1 {\cal T}_n
\left (\begin{array}{c} 1 \\ 0 \end{array}\right ).
\label{a15}
\end{eqnarray}
%--------------------------------------------------------------------
Depending on the defect position the top left element of ${\cal T}_n$,
${{\cal T}_n}{11}$, can attain two values ${\cal T}_+$ or 
${\cal T}_-$, depending whether $\xi_n$ at the cell $n$
is $\xi_+$ or $\xi_-$. The prefactor before the matrix product
equals $\prod_{n=1}^N \xi_n^2=(\lambda+1)^{2(N-K)}(\lambda-1)^{2K}$.
The connection between ${\cal T}_+$
and ${\cal T}_-$ is as follows:
%--------------------------------------------------------------------
\begin{eqnarray}
&&{\cal T}_+= \left (\begin{array}{cc} t & \frac{1}{\beta_L} \\ 
-\frac{1}{\beta_L} &0 \end{array}\right ),\;
{\cal T}_-= \left (\begin{array}{cc} t+\nu & \frac{1}{\beta_L} \\ 
-\frac{1}{\beta_L} &0 \end{array}\right )={\cal T}_++\mbox{diag}(\nu,0),\\
&& z=-1-\lambda-\frac{2}{\beta_L}\left (1+\frac{1}{\eta}\right ),\;
\nu=\frac{4}{(1-\lambda)\eta\beta_L}.
\label{a17b}
\end{eqnarray}
%--------------------------------------------------------------------
At the moment the representation of the characteristic
polynomial (\ref{a15}) is a product of a scalar followed
by the matrix product which can be represented as
%--------------------------------------------------------------------
\begin{eqnarray}\nonumber
&&(1,0)^T \prod_{n=N}^1 {\cal T}_n
\left (\begin{array}{c} 1 \\ 0 \end{array}\right )=
(1,0)^T {{\cal T}_+}^{N-n_K}{\cal T}_- {{\cal T}_+}^{n_K-n_{K-1}-1} {\cal T}_- \cdots  {{\cal T}_-} \times \\
\nonumber
&& \times {{\cal T}_+}^{n_2-n_{1}-1}
 {\cal T}_-{{\cal T}_+}^{n_{1}-1} \left(\begin{array}{c} 1\\0
\end{array} \right )\equiv \\
&& \equiv
\langle N-n_K,n_K-n_{K-1}-1,\cdots,n_2-n_1-1,n_1-1 \rangle .
\label{a18}
\end{eqnarray}
%--------------------------------------------------------------------
The logic behing the last denotation is that we actually have
sequences of ${\cal T}_+$ matrices separated by the defect
matrices ${\cal T}_-$. There numbers in $\langle \cdots \rangle$
are the lengths of the sequences of ${\cal T}_+$. 
With the help of  the Cayley-Hamilton theorem one can find
%--------------------------------------------------------------------
\begin{eqnarray}
{{\cal T}_+}^n= \left (\begin{array}{cc} P_n(z) & \frac{P_{n-1}(z)}{\beta_L} \\ 
-\frac{P_{n-1}(z)}{\beta_L} & -\frac{P_{n-2}(z)}{\beta_L^2} \end{array}\right ),
\end{eqnarray}
where $P_n(z)$ are the polynomials that satisfy the recurrence
relation 
%--------------------------------------------------------------------
\begin{eqnarray}
P_n(z)=zP_{n-1}(z)-\frac{P_{n-2}(z)}{\beta_L^2},
\end{eqnarray}
%--------------------------------------------------------------------
and the first several polynomials have the following explicit
form:
%--------------------------------------------------------------------
\begin{eqnarray}
\nonumber
&&P_0(z)=1,\;P_1(z)=z, \; P_2(z)=z^2-\frac{1}{\beta_L^2},\;
P_3(z)=z^3-\frac{2}{\beta_L^2}z, \\
&& P_4(z)=z^4-\frac{3}{\beta^2_L}z^2+\frac{1}{\beta^4_L}.
\end{eqnarray}
They can be reduced to the second order Chebyshev polynomials
$U_n(x)$ via the following substitution:
\begin{equation}
P_n(z)=\frac{1}{\beta_L^n} U_n(\beta_L z/2).
\end{equation}
One can use the trigonometric form of the Chebyshev polynomials, then
the polynomials $P_n(z)$ can be represented as
\begin{equation}
 P_n(z)=\prod_{k=1}^{n} \left (z-\frac{2}{\beta_L} \cos \frac{k\pi}{n+1}
\right).
\end{equation}

Calculation of the $\langle \cdots \rangle$ term defined in
Eq. (\ref{a18}) is based on the observation that
\begin{eqnarray}
\langle \mbox{diag}({\nu,0})~ {\cal T}_+^n \rangle = \nu P_n(z).
\end{eqnarray}
As a result of that the following is true:
\begin{eqnarray}
\nonumber
&&\langle N-n_K,n_K-n_{K-1}-1,\cdots,n_2-n_1-1,n_1-1 \rangle=\\
\nonumber
&&=\nu P_{n_1-1} \langle N-n_K,n_K-n_{K-1}-1,\cdots,n_2-n_1-1 \rangle+\\
&&+\langle N-n_K,n_K-n_{K-1}-1,\cdots,\cdots,n_2-1 \rangle.
\end{eqnarray}
This operation removes the last interval of $n_1-1$ matrices
${\cal T}_+$ and reduces the size of the $\langle\cdots \rangle$
from $K+1$ to $K$ intervals. Performing this operation 
simultaneously until there is no matrices ${\cal T}_-$ left.

The total expression will hava a binomial sum that is easier
to represent in more explicit form. We take into account the
explicit expression of $\nu$ given in Eq. (\ref{a17b}) and move
it out of the $\langle \cdots \rangle$ brackets. For simplicity
we introduce intervals $\Delta_1=j_1-1$, $\Delta_2=j_2-j_1-1$,
$\ldots$, $\Delta_{K+1}=N-j_K$ that  satisfy $\sum_{l=1}^{K+1}=N-K$.
Finally we obtain
\begin{eqnarray}
\nonumber
&&\langle N-n_K,n_K-n_{K-1}-1,\cdots,n_2-n_1-1,n_1-1 \rangle \equiv \langle \Delta_{K+1},\ldots,\Delta_1 \rangle =\\ 
&&= \nonumber P_N+ 
\nu \sum_{r=1}^{K} P_{\sum_{l=1}^{r} \Delta_l} P_{\sum_{m=r+1}^{K+1}\Delta_m}+\\
\nonumber
&&+\nu^2 \sum_{r=1}^{K(K-1)} P_{\sum_{l=1}^{r} \Delta_l} + \\
&& \cdots \cdots \cdots \nonumber \\
\label{a26}
&& +\nu^{K-1} \sum_{r=1}^K  P_{\Delta_r+\Delta_{r+1}} \prod_{s\neq r,r+1}^K P_{\Delta_s}
+ \nu^K \prod_{r=1}^{K+1} P_{\Delta_r}= \\
\nonumber
&&=(1-\lambda)^{-K} \left \{ (1-\lambda)^{K} P_N+ (1-\lambda)^{K-1}
\left(\frac{4}{\eta\beta_L}\right) \sum_{r=1}^{K} P_{\sum_{l=1}^{r} \Delta_l} P_{\sum_{m=r+1}^{K+1}\Delta_m} + \right .\\
&& \left.  + \cdots + (1-\lambda) \left(\frac{4}{\eta\beta_L}\right)\sum_{r=1}^K  P_{\Delta_r+\Delta_{r+1}} \prod_{s\neq r,r+1}^K P_{\Delta_s}+\prod_{r=1}^{K+1} P_{\Delta_r} \right\}\equiv \\
&&\equiv (-1)^K (\lambda-1) F_{N,K}(z). \nonumber
\label{a27}
\end{eqnarray}

As a result, the whole characteristic polynomial can be written in the
partially 
\begin{eqnarray}
\chi_{N;[j_1,\ldots,j_K]}(\lambda) = (-1)^{K+1} (\lambda+1)^{2(N-K)+1} (\lambda-1)^K F_{N,K}(z),
%&& F_{N,K}=
%+\sum_{n=0}^K (-1)^{K-n}\frac{4^n(\lambda-1)^{K-n}}{(\eta \beta_L)^n}S_{n;K}(t),\\
%&&S_n(t)=\sum_{1\le a_1<\cdots < a_n \le K} P_{j_{a_1}-1}(t)
%\left[ \prod_{m=1}^{n-1} P_{j_{a_{m+1}}-j_{a_m}-1}(t) \right ] 
%P_{N-j_{a_n}}\\
%&& t=-\lambda-1 -\frac{2}{\beta_L}\left (1+\frac{1}{\eta}\right ).
\label{a28}
\end{eqnarray}
where $z$ is given by Eq. (\ref{a17b}) and $F_{N,K}(z)$ is the order
$K$ polynomial given by Eq. (\ref{a28}). 
 
%{\color{red}{
%Several examples of the function $P_n(t)$ can be written
%explicitly (to be removed):
%\begin{eqnarray}
%&&P_1(t)=t-\frac{2}{\beta_L} \cos \frac{\pi}{2}=t,\\
%&&P_2(t)=\left(t-\frac{2}{\beta_L} \cos \frac{\pi}{3}\right)
%\left(t-\frac{2}{\beta_L} \cos \frac{2\pi}{3}\right)=t^2-\frac{1}{\beta^2_L},\\
%&&P_3(t)=\prod_{n=1}^3 \left (t-\frac{2}{\beta_L} \cos \frac{k\pi}{4}
%\right)=t \left (t^2-\frac{2}{\beta^2_L} \right ),\\
%\nonumber
%&&P_4(t)=\prod_{n=1}^4 \left (t-\frac{2}{\beta_L} \cos \frac{k\pi}{5}
%\right)=\left (t^2-\frac{4}{\beta^2_L} \cos^2 \frac{\pi}{5} \right )
%\left (t^2-\frac{4}{\beta^2_L} \cos^2 \frac{2\pi}{5} \right )=\\
%\nonumber
%&&=t^4-\frac{4}{\beta^2_L} t^2\left ( \cos^2 \frac{\pi}{5} + \cos^2 %\frac{2\pi}{5}\right ) + \frac{16}{\beta^4_L}  \cos^2 \frac{\pi}{5} \cos^2 %\frac{2\pi}{5}=\left |\cos \frac{\pi}{5}=\frac{1+\sqrt{5}}{4} \right|=\\
%&&=t^4-\frac{3}{\beta^2_L}t^2+\frac{1}{\beta^4_L} 
%\end{eqnarray}
%}}

Below we consider several examples.
If we have fo fluxons, and/or only asymmetric fluxons, and/or fluxons with even topological charges the is no defects, $K=0$ and the characteristic
polynomial factorizes completely:
\begin{eqnarray}
\!\!\!\!\!
\nonumber
&&\chi_{N}(\lambda) =(-1)^{2N+1} (\lambda+1)^{2N+1}
\prod_{k=1}^{N} \left (z-\frac{2}{\beta_L} \cos \frac{k\pi}{N+1}
\right).
\end{eqnarray}

Suppose there is one fluxon ($K=1$) positioned at the cell $j$.
The auxillary polynomial $F_{N,1}(z)$ will consist of only two
terms. Indeed, we have to expand the matrix product $\langle \cdots \rangle$
only once: $\langle N-j, j-1\rangle=P_N(z)+ \nu P_{N-j}(z) P_{j-1}(z)$.
Consequently, the auxiliary polynomial $F_{N,2}$ has only two terms as well.
Then the resulting 
characteristic polynomial has the following form:
\begin{eqnarray}
\nonumber
&&\chi_{N;[j]}(\lambda)=\\
&&=(\lambda+1)^{2N-1}(\lambda-1)\left [ -(\lambda-1)P_N(t)
+\frac{4}{\eta\beta_L}P_{j-1}(t)P_{N-j}(t) \right ]
\label{a33}
\end{eqnarray}
It is impossible to factorize the term in the square brackets. However,
in the limit of small ladders the auxiliary polynomial can be reduced
to cubic equations. Below we consider these examples.

In the case $N=2$, $j_1\equiv j =2$ the characteristic polynomial 
Eq. (\ref{a33}) reduces to
\begin{eqnarray}
\chi_{2,[2]}=(\lambda+1)^3(\lambda-1)\left [(1-\lambda)P_2(z)
+\frac{4}{\eta\beta_L} P_1(z)P_0(z) \right ] .
%\Rightarrow 
%(\lambda-1)\left(t^2 -\frac{1}{\beta^2_L}\right) - \frac{4}{\eta\beta_L}t=0
\end{eqnarray}
The expression in the square brackets is a cubic polynom which contains
the remaining three eigenvalues. The respective cubic equation can
be rewritten in more compact form
%$$
%t_L\equiv 2 \left[ 1+ \frac{1}{\beta_L}+\frac{1}{\eta\beta_L}\right ]
%\Rightarrow 1-\lambda=t+t_L
%$$
%
\begin{eqnarray}
(z+z_L)\left(z^2-\frac{1}{\beta_L^2} \right)+\frac{4}{\eta\beta_L}z=0,
\;\;\; z_L= 2 \left[ 1+ \frac{1}{\beta_L}+\frac{1}{\eta\beta_L}\right ].
%\Rightarrow \\
%&& \Rightarrow t^3 +t_L t^2 +\frac{1}{\beta_L} \left(\frac{4}{\eta}-
%\frac{1}{\beta_L} \right )t-\frac{t_L}{\beta^2_L}=0
\label{a36}
\end{eqnarray}

For the ladder with $N=3$ and one antisymmetric fluxon placed at the
junction $j=2$ we have
\begin{eqnarray}
\nonumber
&&\chi_{2,[2]}=(\lambda+1)^5 (\lambda-1) \left [
-(\lambda-1)P_3(z)+\frac{4}{\eta\beta_L} P^2_1(z) \right ]= \\
&&=-(\lambda+1)^5 (\lambda-1) z \left [(\lambda-1)\left( z^2 -\frac{2}{\beta^2_L}\right) - \frac{4}{\eta\beta_L}z\right ]=0.
\end{eqnarray}
In addition to the eigenvalues $\lambda=\pm 1$ we have the always negative
eigenvalue $z=0 \Rightarrow$ $\lambda=-[1+2 (1+1/\eta)/\beta_L]$. The remaining three
eigenvalues satisfy
\begin{eqnarray}
&& (z+z_L)\left(z^2-\frac{2}{\beta_L^2} \right)+\frac{4}{\eta\beta_L}z=0.
\end{eqnarray}

Finally, we consider the general case of two defects ($K=2$) placed
at the cells $j_1$ and $j_2$.
In this case we have to expand the product $\langle N-j_1,j_2-j_1-1,j_1-1\rangle$.
This expansion will produce $4$ terms in Eq. (\ref{a27}). The general 
expression for the characteristic polynomial can be written as follows:
\begin{eqnarray}
\chi_{N;[j_1,j_2]}(\lambda)&=&-(\lambda+1)^{2N-3}(\lambda-1)^2 \left \{ (\lambda-1)^2 
P_N(z)+ \right. \nonumber \\
 &+& \left.\frac{4}{\eta \beta_L} (1-\lambda) [P_{j_1-1}(z)P_{N-j_1}(z)
+P_{j_2-1}(z)P_{N-j_2}(z) ] +  \right . \nonumber \\
 &+& \left. \frac{16}{(\eta \beta_L)^2} P_{j_1-1}(z)
P_{j_2-j_1-1}(z) P_{j_2-j_1}(z)\right \} .
\end{eqnarray}

%--------------------------------------------------------------------
%\begin{eqnarray}
%&&    \boxed{\det(M - \lambda I) = 2^{2-N} (-\lambda - 1)^{2N+1} \cdot \left[ (W^2 - \frac{1}{\beta_L^2}) S_{N-1} + \frac{2}{\beta_L^2} W S_{N-2} \right ]}
%    \label{eq:matrix},\\
%&& \nonumber  S_N=\sum_{k=1 ; k \text{odd}}^{N} \binom{N}{k} (-W)^{N-k} (W^2 - \frac{4}{\beta_L^2})^{\frac{k-1}{2}}=|k=2m-1|=\\
%    &=&\sum_{m=1}^{N'} \binom{N}{2m-1}
%    (-W)^{N-2m+1} (W^2 - \frac{4}{\beta_L^2})^{m-1}, \\
%    && N'=(N-1)/2 [N\;\text{odd}];
%N'=N/2-1 [N\; \text{even}] ,\\
%&&    W =\lambda+1+2(\frac{1}{\beta_L \eta}+\frac{1}{\beta_L}) 
%\end{eqnarray}
%where $W =\lambda+1+2(\frac{1}{\beta_L \eta}+\frac{1}{\beta_L})$.
%\begin{eqnarray}
%\nonumber  S_N&=&\sum_{k=1 ; k \text{odd}}^{N} \binom{N}{k} (-W)^{N-k} (W^2 - \frac{4}{\beta_L^2})^{\frac{k-1}{2}}=|k=2m-1|=\\
%    &=&\sum_{m=1}^{N'} \binom{N}{2m-1}
%    (-W)^{N-2m+1} (W^2 - \frac{4}{\beta_L^2})^{m-1}, \\
%    && N'=(N-1)/2 [N\;\text{odd}];
%N'=N/2-1 [N\; \text{even}]   . 
%\end{eqnarray}

%\bibliographystyle{unsrt}
%\bibliographystyle{elsarticle-num}
%\bibliography{/home/yzolo/TEX/BIB/fb,/home/yzolo/TEX/BIB/jj,/home/yzolo/TEX/BIB/sc,/home/yzolo/TEX/BIB/books,/home/yzolo/TEX/BIB/el,/home/yzolo/TEX/BIB/bec,/home/yzolo/TEX/BIB/math,/home/yzolo/TEX/BIB/db,/home/yzolo/TEX/BIB/magn,/home/yzolo/TEX/BIB/int}

\begin{thebibliography}{10}
\expandafter\ifx\csname url\endcsname\relax
  \def\url#1{\texttt{#1}}\fi
\expandafter\ifx\csname urlprefix\endcsname\relax\def\urlprefix{URL }\fi
\expandafter\ifx\csname href\endcsname\relax
  \def\href#1#2{#2} \def\path#1{#1}\fi

\bibitem{tmo00prl}
E.~Tr\'ias, J.~J. Mazo, T.~P. Orlando, Discrete breathers in nonlinear
  lattices: Experimental detection in a josephson array, Phys. Rev. Lett.
  84~(4) (2000) 741--744.
\newblock \href {https://doi.org/10.1103/PhysRevLett.84.741}
  {\path{doi:10.1103/PhysRevLett.84.741}}.

\bibitem{baufz00prl}
P.~Binder, D.~Abraimov, A.~V. Ustinov, S.~Flach, Y.~Zolotaryuk, Observation of
  breathers in josephson ladders, Phys. Rev. Lett. 84~(4) (2000) 745--748.

\bibitem{fs99jpc}
S.~Flach, M.~Spicci, Rotobreather dynamics in underdamped josephson junction
  ladders, J. Phys.: Cond. Matt. 11~(1) (1999) 321--334.

\bibitem{mffzp01pre}
A.~E. Miroshnichenko, S.~Flach, M.~V. Fistul, Y.~Zolotaryuk, J.~B. Page,
  Breathers in josephson junction ladders: Resonances and electromagnetic wave
  spectroscopy, Phys. Rev. E 64~(6) (2001) 066601.

\bibitem{bz22jpc}
D.~Bukatova, Y.~Zolotaryuk, Flat and almost flat bands in the
  quasi-one-dimensional josephson junction array, J. Phys.: Condens. Matte 34
  (2022) 175402.

\bibitem{tmo02prb}
E.~Tr\'\i{}as, J.~J. Mazo, T.~P. Orlando, Interactions between josephson
  vortices and breathers, Phys. Rev. B 65 (2002) 054517.

\bibitem{wzso96pd}
S.~Watanabe, H.~S.~J. van~der Zant, S.~H. Strogatz, T.~P. Orlando, Dynamics of
  circular arrays of josephson junctions and the discrete sine-gordon
  equations, Physica D 97 (1996) 429--470.

\bibitem{u98pd}
A.~V. Ustinov, Solitons in josephson junctions, Physica D 123~(1-4) (1998)
  315--329.

\bibitem{mff95prbII}
J.~J. Mazo, F.~Falo, L.~M. Flor\'{\i}a,
  \href{https://link.aps.org/doi/10.1103/PhysRevB.52.10433}{Josephson junction
  ladders: Ground state and relaxation phenomena}, Phys. Rev. B 52 (1995)
  10433--10440.
\newblock \href {https://doi.org/10.1103/PhysRevB.52.10433}
  {\path{doi:10.1103/PhysRevB.52.10433}}.
\newline\urlprefix\url{https://link.aps.org/doi/10.1103/PhysRevB.52.10433}

\bibitem{gfpg96pla}
G.~Grimaldi, G.~Filatrella, S.~Pace, U.~Gambardella, Flux distribution and
  critical currents in a one-dimensional row of a josephson junction square
  lattice, Phys. Lett. A 223 (1996) 463--469.

\bibitem{psz25arx}
A.~Prykhodko, I.~O. Starodub, Y.~Zolotaryuk, Compact stationary fluxons in the
  josephson junction ladder (2025).
\newblock \href {http://arxiv.org/abs/arXiv:2511.19215 [cond-mat.supr-con]}
  {\path{arXiv:arXiv:2511.19215 [cond-mat.supr-con]}}, \href
  {https://doi.org/https://doi.org/10.48550/arXiv.2511.1921}
  {\path{doi:https://doi.org/10.48550/arXiv.2511.1921}}.

\bibitem{aks10prl}
F.~K. Abdullaev, P.~G. Kevrekidis, M.~Salerno,
  \href{https://link.aps.org/doi/10.1103/PhysRevLett.105.113901}{Compactons in
  nonlinear schr\"odinger lattices with strong nonlinearity management}, Phys.
  Rev. Lett. 105 (2010) 113901.
\newblock \href {https://doi.org/10.1103/PhysRevLett.105.113901}
  {\path{doi:10.1103/PhysRevLett.105.113901}}.
\newline\urlprefix\url{https://link.aps.org/doi/10.1103/PhysRevLett.105.113901}

\bibitem{ssmjsvs20pra}
M.~G. Stojanovi{\ifmmode \acute{c}\else \'{c}\fi{}}, M.~S. Krasi{\ifmmode
  \acute{c}\else \'{c}\fi{}}, A.~Maluckov, M.~Johansson, I.~A. Salinas, R.~A.
  Vicencio, M.~Stepi{\ifmmode \acute{c}\else \'{c}\fi{}},
  \href{https://link.aps.org/doi/10.1103/PhysRevA.102.023532}{Localized modes
  in linear and nonlinear octagonal-diamond lattices with two flat bands},
  Phys. Rev. A 102 (2020) 023532.
\newblock \href {https://doi.org/10.1103/PhysRevA.102.023532}
  {\path{doi:10.1103/PhysRevA.102.023532}}.
\newline\urlprefix\url{https://link.aps.org/doi/10.1103/PhysRevA.102.023532}

\bibitem{barone82}
A.~Barone, G.~Paterno, Physics and Applications of the Josephson Effect, Wiley,
  New York, 1982.

\bibitem{fmamfm98pd}
L.~M. Floria, J.~L. Marin, S.~Aubry, P.~J. Martinez, F.~Falo, J.~J. Mazo,
  Josephson-junction ladder: a benchmark for nonlinear concepts, Physica D 113
  (1998) 387.

\bibitem{kittel}
C.~Kittel, Introduction To Solid State Physics, John Wiley and Sons, New York,
  2005.

\bibitem{bk03}
O.~M. Braun, Y.~S. Kivshar, The Frenkel-Kontorova Model: Concepts, Methods, and
  Applications, Springer, Berlin, 2003.

\end{thebibliography}

\end{document}